\documentclass[aps,superscriptaddress,twocolumn]{revtex4-2}
\usepackage[T1]{fontenc}
\usepackage{babel}[english]
\usepackage{amsmath, amsfonts, mathtools, mathrsfs}
\usepackage{graphicx, float}
\usepackage[dvipsnames]{xcolor}
\usepackage{physics}
\usepackage{mleftright}
\usepackage{soul}
\usepackage{tikz, circuitikz}
\usepackage{hyperref}
\usetikzlibrary{calc, arrows.meta}
\definecolor{darkred}{cmyk}{0, 1, 1, 0.2}
\definecolor{darkblue}{cmyk}{1, 1, 0, 0}

\definecolor{ahs-color}{rgb}{0,0.7,0.8}
\definecolor{tp-color}{rgb}{0.2,0.6,0}
\definecolor{fc-color}{rgb}{0.7,0,0.5}

\begin{document}

\title{Geodesic-based optimal control for leakage suppression in superconducting qubits}

\author{Adonai H. da Silva}
\email{Adonai.HilariodaSilva@vtt.fi}
\affiliation{Quantum algorithms and software, VTT Technical Research Centre of Finland Ltd, Tietotie 3, 02150 Espoo, Finland}

\author{Tanmoy Pandit}
\affiliation{Quantum algorithms and software, VTT Technical Research Centre of Finland Ltd, Tietotie 3, 02150 Espoo, Finland}

\author{Francesco Cosco}
\affiliation{Quantum algorithms and software, VTT Technical Research Centre of Finland Ltd, Tietotie 3, 02150 Espoo, Finland}

\begin{abstract}
% \st{Leakage errors constitute a one of the major challenge for fault-tolerant quantum computing with superconducting qubits, where weak anharmonicity can populate non-computational states during gate operations.}
Superconducting qubits are often based on the first two energy levels of a physical anharmonic LC oscillator.
Since the anharmonicity of such systems is usually much lower than the involved energy gaps, population can easily leak to higher energy states. This is what characterizes leakage errors, and it can have multiple sources: from coherent error induced by the control platform to incoherent thermal dissipation.
% \st{We develop a geodesic-based optimal control framework that formulates quantum gate synthesis as a sub-Riemannian optimization problem on the unitary manifold while incorporating smooth envelope constraints for experimental realizability. Using a realistic model of two transmon qubits coupled through a tunable coupler, we derive an effective two-qutrit Hamiltonian and investigate leakage-aware gate design in the presence of thermal relaxation.}
In this work, we adapt an optimal control method based on sub-Riemannian geodesic search in the unitary group for a system consisting of two superconducting qubits with leakage levels.
%\st{For single-qubit $R_x(\pi/2)$ gates, the optimized pulses outperform Gaussian and DRAG protocols in the short-gate regime, simultaneously reducing leakage and seepage while achieving fidelities close to the thermal-noise limit. In particular, pulse durations as short as $17\,\mathrm{ns}$ yield leakage rates below $10^{-5}$ and near-optimal gate performance.}
We use the $R_x(\pi/2)$ rotation for single-qubit operation analysis, and the $i$SWAP operation for two-qubit analysis. For the former, we find considerable performance improvement in average gate fidelity as well as in the reduction of leakage errors when comparing the obtained optimal pulses with Gaussian and DRAG pulses, especially for short duration pulses. For the $i$SWAP gate, although leakage error remains roughly the same, we manage to obtain considerably higher fidelities with the cost of requiring a slightly longer time evolution.
%\st{The method naturally extends to two-qubit operations such as the iSWAP gate, demonstrating the applicability of geometric optimal control for multi-qubit superconducting architectures. Our results establish sub-Riemannian geodesic control as a promising approach for leakage-aware quantum gate engineering in quantum processors.}
\end{abstract}

\maketitle

% ========================================================================================================== %
% Paper begins
% ========================================================================================================== %

\section{Introduction}
%\st{Quantum computing has emerged as one of the most promising paradigms for information processing, offering the potential to solve certain computational problems exponentially faster than classical algorithms %\cite{Nielsen2010,Shor1997,preskill2018quantum,grover1997quantum}.
%Among the various hardware platforms currently under development,}

With the advent of quantum computing, several different physical platforms were considered for building a quantum computer, each with their own advantages and limitations. Some examples are trapped ions~\cite{bruzewicz2019trapped, zalivako2025quantum}, neutral atoms~\cite{henriet2020quantum, bluvstein2026fault}, photonics~\cite{kok2007linear, aghaee2025scaling}, nitrogen-vacancy centers in diamonds~\cite{childress2013diamond, khazen2023nv}, and quantum dots~\cite{burkard2000spintronics, kloeffel2013prospects}. In particular, one of the most promising ones are superconducting circuits,~\cite{Koch2007, marxer2023IQM} owing to their operational speed, 
%\fc{Comment: maybe remove scalability and point at operational speed, scaling up superconducting qubits has still its own challenges}
compatibility with microfabrication technologies~\cite{marxer2026above, conchello2026plaquette}, and successful experimental progress toward the implementation of quantum error correction (QEC) codes \cite{Krantz2019,Kjaergaard2020,Blais2021,Fowler2012,GoogleQEC2023,IBMQEC2024}.

Common to all platforms is the necessity of developing control tools for delivering operations on the physical qubits, as well as minimizing effects caused by the presence of noise. It is known that in order to have reliable QEC for fault tolerant quantum computers it is necessary to fulfill a certain threshold in the fidelity of quantum operations.~\cite{nielsen2010, gaitan2008} For superconducting qubits specifically, one relevant source of errors that hinders quantum fidelity is leakage noise.~\cite{Suchara2015, Aliferis2007, Wood2024} This type of noise is characterized by a depopulation of the computational subspace. It happens because modern transmon technologies use the first two energy levels of a quantum anharmonic oscillator to construct the qubit, however, imprecision in the external control fields and incoherent processes unavoidably cause some population to leak to these higher energy states. This is particularly detrimental to QEC since it cannot be described in terms of Pauli operators. Thus there is considerable interest in methods of quantum control that address this problem. A natural approach to suppressing general decoherence is to execute quantum gates as fast as possible using time-optimal control protocols.~\cite{jirari2020time, jirari2026time} However, while this approach is useful for reducing noise originating from the qubit interacting with an external environment, it can induce considerable leakage noise because the resulting control fields are usually composed of fast turn-on and turn-off operations, which broaden the control spectrum in the frequency domain. As a consequence, considerable amplitude values appear in the frequencies corresponding to higher energy transitions.

Inspired by geometric-based methods for treating quantum complexity~\cite{nielsengeo1, nielsengeo2, browngeo} and even time-optimality~\cite{lloydbrach}, we use the method first introduced in ~\cite{morazotti2024optimized, da2025monte} and adapt it to a realistic two-qubit coupled via a tunable coupler system that takes into account leakage effects during the dynamics. More specifically, the unitary evolution of quantum systems can be interpreted as a trajectory on the Lie group corresponding to that system's dynamics. Similarly, quantum gates correspond to target points on this manifold and the system Hamiltonian is the generator of trajectories, meaning it lives in the associated Lie algebra. Since external drives cannot manipulate all degrees of freedom, the control Hamiltonian is restricted to the projection of the tangent vector onto the controllable subspace.

We benchmark this approach using the single-qubit $R_x(\pi/2)$ gate and the two-qubit $i$SWAP gate. Swap interactions are particularly relevant for superconducting platforms because of the low degree of connectivity among qubits, which are generally in fixed positions in the quantum chip, limiting the applicability of some QEC codes without an efficient way of performing swap operations ~\cite{mcewen2021removing, miao2023overcoming}. We remark the existence of several alternative methods for quantum control. For this work in particular, we compare the pulses optimized by means of the sub-Riemannian geodesic search with derivative removal by adiabatic gate (DRAG) pulses~\cite{Motzoi2009, theis2018counteracting} and show that for short duration pulses, the method presented in this work shows a better performance for both: increasing gate fidelity and minimizing leakage error rate.

This paper is organized as follows. In Section~\ref{sec:optimal_control} we make a quick review of the geodesic optimal control formalism from ~\cite{morazotti2024optimized, da2025monte} and introduce the modifications required for realistic pulse synthesis in superconducting platforms. In Section~\ref{sec:physical_model} we discuss the physical model and obtain the effective Hamiltonian describing the architecture consisting of two qubits coupled directly and via a tunable coupler, and explain how quantities such as fidelity, leakage, and seepage~\cite{Wood2024} are quantified in this work. In Section~\ref{sec:results} we show the results of the simulated one- and two-qubit operations, and compare the performance with other well known existing control methods in the literature. Finally, in Section~\ref{sec:conclusion} we summarize the main findings and discuss possible future research  directions. 

\section{Geodesic-based optimal control}\label{sec:optimal_control}

In this section we briefly review the optimal control method introduced in~\cite{morazotti2024optimized, da2025monte}, and lay out some important modifications for dealing with leakage noise processes. In general, a geodesic is a curve in a Riemannian manifold that connects two points through the shortest path.~\cite{robbin2022} One way to define them is through the functional
\begin{align}\label{eq:Length_functional}
    \mathscr{L}(\gamma) \equiv \int_{t_0}^{t_f} \mathrm{d}t \, \sqrt{\langle{\dot{\gamma}(t), \dot{\gamma}(t)}\rangle}.
\end{align}
$t_0$ and $t_f$ parametrize the initial and final points, $\dot{\gamma}(t)$ is the tangent vector to a curve $\gamma(t)$, interpreted as a velocity vector, and $\braket{\cdot}{\cdot}$ is an inner product, that will naturally induce some metric tensor. The geodesic is then the path $\gamma(t)$ for which $\mathscr{L}$ is minimal. If $\dot{\gamma}(t)$ lives in the entire tangent space, we call the optimal $\gamma(t)$ a Riemannian geodesic. If the $\dot{\gamma}(t)$ are restricted to only a subspace of the tangent space, $\gamma(t)$ is called a sub-Riemannian geodesic.~\cite{agrachev2020} As it will be shown later, the geodesics we search for in the context this work are sub-Riemannian.

In general, we consider a quantum system whose time evolution can be described by elements of a Lie group $\mathrm{G}$. For a system consisting of $n$ qudits ($d$-dimensional generalization of qubits) evolving unitarily, we have $\mathrm{G}=\mathrm{SU}(d^n)$. Such a group is a Riemannian manifold, where the tangent space is given by the Lie algebra $\mathfrak{su}(d^n)$ over the identity element. So we can treat unitary time evolution operators $U(t)$, parameterized by time, as curves connecting the identity operator ($U(0) \equiv \mathbb{I}$) to some desired unitary operation $U(\tau)$. The object that acts as tangent vector is the system's entire Hamiltonian $H(t)$. This is a direct consequence of the Schr\"odinger equation in operator form
\begin{align}\label{eq:schrodinger_operational}
    \frac{\mathrm{d}U}{\mathrm{d}t} = -i H(t) U(t),
\end{align}
where we are using natural units of $\hbar = 1$, and will use for the remainder of this paper. If we were able to completely tune and control the entire $H(t)$ directly, then the optimal path $U(t)$ would be a Riemannian geodesic. However, in the most general scenario, the entire system Hamiltonian can be divided as follows
\begin{align}
    H(t) = H_0 + H_\mathrm{int}(t) + H_\mathrm{c}(t).
\end{align}
$H_0$ describing the intrinsic eigenenergies of the system, $H_\mathrm{int}(t)$ accounting for any interaction terms between different parts of the system, and $H_\mathrm{c}(t)$ corresponds to an external control Hamiltonian. $H_\mathrm{c}(t)$ is usually the only part of $H(t)$ that we can smoothly tune, and the part we want to optimize. So analogously to Eq.~(\ref{eq:Length_functional}), we can define an energy functional
\begin{align}\label{eq:energy_functional}
    \mathscr{E}(H) \equiv \frac{1}{2} \int_0^\tau \mathrm{d}t \, \langle{H_\mathrm{c}(t), H_\mathrm{c}(t)}\rangle. 
\end{align}
Notice that only components in the subspace spanned by $H_\mathrm{c}(t)$ contribute to the cost functional. We say then that unitary evolution operators which minimize Eq.~(\ref{eq:energy_functional}) are sub-Riemannian geodesics. Here we define the inner product as the normalized trace product, that is,
\begin{align}
    \langle{x, y}\rangle \equiv \frac{1}{d_\mathrm{c}} \tr(x \cdot y),
\end{align}
where $d_\mathrm{c}$ denotes the dimension of the control subspace, and $x, y$ are vectors.

By considering that the control operations form a subspace we are effectively writing the Lie algebra as direct sum of vector spaces $\mathfrak{g} = \mathcal{V} \oplus \mathcal{V}^\perp$, where $\mathcal{V}$, usually called a \textit{distribution} in the context of sub-Riemannian geometry, corresponds to the subspace spanned by the control operators, while $\mathcal{V}^\perp$ corresponds to all operators we are not able to directly control.

Imposing minimization of Eq.~(\ref{eq:energy_functional}) is not enough though, because it does not contain the constraint given by the Schr\"odinger equation, Eq.~(\ref{eq:schrodinger_operational}). We then must introduce a costate $\Lambda(t)$ that enforces that Eq.~(\ref{eq:schrodinger_operational}) is satisfied for all times. This results in a functional given by
\begin{equation}
\begin{aligned}
    \mathscr{J}(H,U,\Lambda) &\equiv \mathscr{E}(H) \\ &\quad + \int_0^\tau \mathrm{d}t \, \left\langle \Lambda(t),  \frac{\mathrm{d}U}{\mathrm{d}t} + i H(t) U(t) \right\rangle.
\end{aligned}
\end{equation}
The costate is defined as $\Lambda(t) \equiv \sum_k \lambda_k(t) \zeta_k$, where the index $k$ runs from $1$ to the dimension of the entire algebra, $\lambda_k(t)$ are real functions, and the elements $\zeta_k$ form a basis for the algebra. For a system of $n$ qudits, $\zeta_k$ can be taken to be $\sigma_{\mu,1} \otimes \sigma_{\mu,2} \otimes \cdots \otimes \sigma_{\mu,n}$, where each $\sigma_{\mu,j}$ is the $\mu$-th generalized Pauli matrix for the $j$-th qudit.
Following the methodology shown in ~\cite{morazotti2024optimized, da2025monte}, we consider fixed points for $t=0$ and $t=\tau$, and require that variations of $H$, $U$, and $\Lambda$ do not change the functional around the minimum of $\mathscr{J}$. This process is based on Pontryagin's Maximum principle, which can be considered a generalization of Lagrangian calculus.~\cite{pontryagin1987, boscain2021introduction} This process results in the geodesic equation, given by
\begin{align}\label{eq:base_geodesiceq}
    \frac{\mathrm{d}U}{\mathrm{d}t} = -i \left[ H_\mathrm{nc}(t) + \mathrm{P}_\mathcal{V} \mleft[ U(t) \Lambda(0) U^\dagger(t) \mright] \right] U(t).
\end{align}
$H_\mathrm{nc}(t)$ corresponds to the uncontrollable Hamiltonian, that is, $H_0 + H_\mathrm{int}(t)$ rewritten in some conventional frame, usually the interaction picture, where we consider the frame co-rotating with the qubits' frequencies. $\mathrm{P}_\mathcal{V}$ corresponds to the projection onto the distribution $\mathcal{V}$. Notice that the equation depends only on the initial value of the costate $\Lambda(0)$. Once an initial condition is set, the equation can be integrated numerically from $0$ to $\tau$, resulting in some two-qubit operation $U(\tau)$. The objective is therefore determining what initial condition $\Lambda(0)$ will result in the specific desired operation. For that, we follow the shooting method laid out in ~\cite{da2025monte}, based on guessing a sufficient number of random initial costates, and using a gradient ascent function for maximizing the process fidelity using the best initial condition as an initial guess.

Equation~(\ref{eq:base_geodesiceq}) is the non-linear differential equation solved in ~\cite{morazotti2024optimized, da2025monte}. Here we introduce one further constraint, which is that the control Hamiltonian must continuously rise and fall to zero in the beginning and ending of the pulse to make it smooth and reasonable for experimental implementation. In the previous works this was not considered, resulting in Hamiltonians that would require fast turn on and off operations when implemented in realistic experiments. Abrupt rise and fall in control functions cause a broadening of the spectrum in the frequency domain, inducing  leakage errors in the system. To prevent this, we introduce an envelope function $S(t)$ multiplying the control function $\mathrm{P}_\mathcal{V} \left[ U(t) \Lambda(0) U^\dagger(t) \right]$, so the sub-Riemannian geodesic equation is altered to
\begin{align}\label{eq:geodesiceq_this_work}
    \frac{\mathrm{d}U}{\mathrm{d}t} = -i \left[ H_\mathrm{nc}(t) + S(t) \mathrm{P}_\mathcal{V} \mleft[ U(t) \Lambda(0) U^\dagger(t) \mright] \right] U(t).
\end{align}
There are several choices for the envelope $S(t)$, such as a sinusoidal rise and fall, however, this requires a piece-wise definition for $S(t)$. A simpler choice is a product of two smoothed Heaviside step functions. This can be achieved by means of the sigmoid function, defined as
\begin{align}
    \theta_r(t) \equiv \frac{1}{1+e^{-rt}},
\end{align}
where $r$ is a relaxation factor. By setting an offset time $t_\mathrm{off}$ such that the value is very close to zero at $t=0$ and at $t = \tau$, we define the envelope function as
\begin{align}\label{eq:envelope}
    S(t) \equiv \theta_r(t-t_\mathrm{off}) \theta_r(\tau - t - t_\mathrm{off}).
\end{align}
%\tp{ Comment: \[ S(\tau) = \theta_r\!\left(\tau-t_{\mathrm{off}}\right) \theta_r\!\left(t_{\mathrm{off}}\right). \] For $\tau > t_{\mathrm{off}}$ and $r\,t_{\mathrm{off}} \gg 1$, both logistic factors are close to unity, yielding \[ S(\tau) \approx 1, \] rather than the expected $S(\tau) \approx 0$. Hence, Eq.~(10) does not produce a pulse envelope that smoothly switches off at the final time. I believe there is a sign error in the definition of the envelope. The issue would be resolved if the second sigmoid were replaced by \[ \theta_r\!\left(\tau - t - t_{\mathrm{off}}\right), \] rather than \[ \theta_r\!\left(\tau - t + t_{\mathrm{off}}\right). \], no?? }
For the purposes of this work, we chose the relaxation factor to be $r = 2.4$, and the offset $t_\mathrm{off} = 6$ ns. This choice roughly sets a buffer time of approximately $4$ ns in the beginning and end of the pulse. During these intervals, the amplitude values are less than $0.1$ MHz.

\section{Physical model}\label{sec:physical_model}

We next illustrate the application of the geodesic optimal control formalism to realistic superconducting qubit architectures exhibiting leakage errors. This section is divided as follows. In Sec.~\ref{subsec:two-qubit_system}, we detail the physical configuration of the system consisting of two superconducting qubits, the tunable coupler, and the individual capacitive control lines. In Sec.~\ref{subsec:effective_Hamiltonian}, we obtain an effective Hamiltonian for the two-qubit subspace within the appropriate approximations. In section~\ref{subsec:2qt-algebra}, we lay out the algebra and control subspace for the two-qubit system, using the concepts introduced in section~\ref{sec:optimal_control}. In Sec.~\ref{subsec:leakage_and_seepage} we briefly explain how leakage and seepage rates are calculated, and how they change the calculation for average gate fidelity.

\subsection{Superconducting two-qubit system}\label{subsec:two-qubit_system}

We model a single-qubit as a Duffing oscillator \cite{Krantz2019}, which is described by the Hamiltonian
\begin{align}
    H_\mathrm{qb} = \omega_{01} a^\dagger a + \frac{\alpha}{2} a^\dagger a^\dagger a a,
\end{align}
where $\omega_{jk}$ is the angular transition frequency between levels $\ket{j}$ and $\ket{k}$, $\alpha \equiv \omega_{12} - \omega_{01}$ is the anharmonicity, $a^\dagger$ and $a$ are the usual ladder operators inducing excitation and de-excitation, respectively.
%, in the natural units with $\hbar = 1$.

The entire system is composed by a pair of qubits and a tunable coupler, which for practical means, can also be considered to be a Duffing oscillator. The native Hamiltonian for the entire system will thus be
\begin{equation}\label{eq:H0}
\begin{aligned}
    H_0 &= \omega_c a_c^\dagger a_c + \frac{\alpha_c}{2} a_c^\dagger a_c^\dagger a_c a_c \\ &\quad + \sum_{k=1}^2 \omega_k a_k^\dagger a_k + \frac{\alpha_k}{2} a_k^\dagger a_k^\dagger a_k a_k,
\end{aligned}
\end{equation}
where now we use $\omega_k$ to specifically denote the $\omega_{01}$ frequency of the $k$-th qubit.

We consider that each qubit is capacitively coupled to the tunable coupler, as well as directly between them. A capacitive coupling between qubits $j$ and $k$ is described by the Hamiltonian
\begin{align}
    H_\mathrm{cc} = g (-i) \left( a_j - a_j^\dagger \right) (-i) \left( a_k - a_k^\dagger \right),
\end{align}
where $g$ is the coupling strength. With this, the interaction Hamiltonian among the two qubits and the coupler can be written as
\begin{equation}\label{eq:Hint}
\begin{aligned}
    H_\mathrm{int} &= - \sum_{k=1}^2 g_k \left( a_k a_c - a_k a_c^\dagger - a_k^\dagger a_c + a_k^\dagger a_c^\dagger \right) \\ &\quad - g_{12} \left( a_1 a_2 - a_1 a_2^\dagger - a_1^\dagger a_2 + a_1^\dagger a_2^\dagger \right),
\end{aligned}
\end{equation}
where $g_k$ denotes the coupling strength between the coupler and the $k$-th qubit, and $g_{12}$ the direct qubit-qubit coupling strength.

Lastly, we consider each qubit to be coupled to an external AC source for control, which can be described by the drive Hamiltonian
\begin{align}\label{eq:HAC}
    H_\mathrm{d}(t) = \sum_{k=1}^2 -i c_k V_{\mathrm{d},k}(t) \left( a_k - a_k^\dagger \right),
\end{align}
where $c_k$ is the coupling strength, and $V_{\mathrm{d},k}(t)$ is the time dependent applied voltage, described by
\begin{equation}\label{eq:Vdk_original}
\begin{aligned}
    V_{\mathrm{d},k}(t) &= V_{I,k}(t) \cos(\omega_{\mathrm{d},k} t + \phi_k) \\ &\quad + V_{Q,k}(t) \sin(\omega_{\mathrm{d},k} t + \phi_k),
\end{aligned}
\end{equation}
where $V_{I,k}(t)$ and $V_{Q,k}(t)$ are the in-phase and quadrature signal modulations, respectively, $\omega_{\mathrm{d},k}$ is the control drive frequency, and $\phi_k$ is a phase.

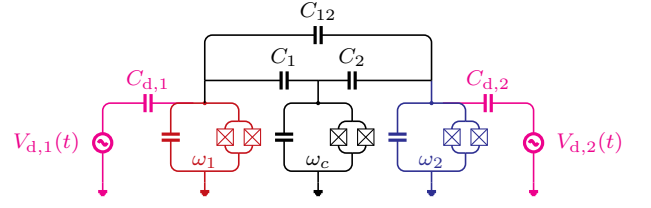
\begin{figure}[]
  \centering
  \begin{tikzpicture}[scale=0.3,
    font=\footnotesize,
    rounded corners=1mm,
    % 1. Sets line width for ALL regular wires and paths globally
    %every path/.style={line width=0.55pt}, 
    jj/.style={
        rectangle, 
        draw, 
        minimum size=0.6mm,
        rounded corners=0mm,      
        path picture={
            \draw (path picture bounding box.north west) -- (path picture bounding box.south east);
            \draw (path picture bounding box.south west) -- (path picture bounding box.north east);
        }
    }
]

% ---------------------------------------------------
% GLOBAL CIRCUITIKZ SETTINGS
% ---------------------------------------------------
\ctikzset{
    bipoles/length=0.4cm,
    nodes width=0.05,
    % 2. Scales the component symbols (capacitors, AC sources) to match 0.55pt
    %bipoles/thickness=1.8
}

% ---------------------------------------------------
% LEFT CIRCUIT (Qubit 1)
% ---------------------------------------------------
% Drive line (left) - Combined into a continuous path
    \draw[color=magenta, line width=0.6pt] (-5.5, 3.5) to[C, l_=$C_{\mathrm{d},1}$] (-9.5, 3.5) to[sV, l_=$V_{\mathrm{d},1}(t)$] (-9.5, 0) node[ground] {};
\begin{scope}[darkred]
    
    % Main Loop - Combined into a continuous path for rounded corners
    \draw[line width=0.6pt] (-3.5, 1.3) -- (-3.5, 0.5) -- (-6.5, 0.5) to[C, color=darkred] (-6.5, 3.5) -- (-3.5, 3.5) -- (-3.5, 2.7);
    
    % Ground stem
    \draw[line width=0.6pt] (-5, 0.5) -- (-5, 0) node[ground] {};
    \node at (-5, 0.9) {$\omega_1$};
    
    % SQUID Loop - Widened and heightened
    \node[jj] (JJ1_L) at (-4.1, 2.0) {};
    \node[jj] (JJ1_R) at (-2.9, 2.0) {};
    
    \draw[line width=0.6pt] (-3.5, 1.3) -- (-4.1, 1.3) -- (JJ1_L.south); 
    \draw[line width=0.6pt] (JJ1_L.north) -- (-4.1, 2.7) -- (-3.5, 2.7);
    \draw[line width=0.6pt] (-3.5, 1.3) -- (-2.9, 1.3) -- (JJ1_R.south); 
    \draw[line width=0.6pt] (JJ1_R.north) -- (-2.9, 2.7) -- (-3.5, 2.7);
    
    % Upward connection
    \node[circ, color=darkred] at (-5, 3.5) {};
\end{scope}
    \draw[line width=0.6pt] (-5, 3.5) -- (-5, 4.5) coordinate (C1_top);

% ---------------------------------------------------
% CENTER CIRCUIT (Coupler)
% ---------------------------------------------------
\begin{scope}[]
    % Main Loop
    \draw[line width=0.6pt] (1.5, 1.3) -- (1.5, 0.5) -- (-1.5, 0.5) to[C] (-1.5, 3.5) -- (1.5, 3.5) -- (1.5, 2.7);
    
    % Ground stem
    \draw[line width=0.6pt] (0, 0.5) -- (0, 0) node[ground] {};
    \node at (0, 0.9) {$\omega_c$};
    
    % SQUID Loop - Widened and heightened
    \node[jj] (JJC_L) at (0.9, 2.0) {};
    \node[jj] (JJC_R) at (2.1, 2.0) {};
    
    \draw[line width=0.6pt] (1.5, 1.3) -- (0.9, 1.3) -- (JJC_L.south); 
    \draw[line width=0.6pt] (JJC_L.north) -- (0.9, 2.7) -- (1.5, 2.7);
    \draw[line width=0.6pt] (1.5, 1.3) -- (2.1, 1.3) -- (JJC_R.south); 
    \draw[line width=0.6pt] (JJC_R.north) -- (2.1, 2.7) -- (1.5, 2.7);
    
    % Upward connection
    \node[circ] at (0, 3.5) {};
    \draw[line width=0.6pt] (0, 3.5) -- (0, 4.5) coordinate (Cc_top);
\end{scope}

% ---------------------------------------------------
% RIGHT CIRCUIT (Qubit 2)
% ---------------------------------------------------
% Drive line (right)
    \draw[color=magenta, line width=0.6pt] (5.5, 3.5) to[C, l=$C_{\mathrm{d},2}$] (9.5, 3.5) to[sV, l=$V_{\mathrm{d},2}(t)$] (9.5, 0) node[ground] {};
\begin{scope}[darkblue]
    % Main Loop
    \draw[line width=0.6pt] (6.5, 1.3) -- (6.5, 0.5) -- (3.5, 0.5) to[C, color=darkblue] (3.5, 3.5) -- (6.5, 3.5) -- (6.5, 2.7);
    
    % Ground stem
    \draw[line width=0.6pt] (5, 0.5) -- (5, 0) node[ground] {};
    \node at (5, 0.9) {$\omega_2$};
    
    % SQUID Loop - Widened and heightened
    \node[jj] (JJ2_L) at (5.9, 2.0) {};
    \node[jj] (JJ2_R) at (7.1, 2.0) {};
    
    \draw[line width=0.6pt] (6.5, 1.3) -- (5.9, 1.3) -- (JJ2_L.south); 
    \draw[line width=0.6pt] (JJ2_L.north) -- (5.9, 2.7) -- (6.5, 2.7);
    \draw[line width=0.6pt] (6.5, 1.3) -- (7.1, 1.3) -- (JJ2_R.south); 
    \draw[line width=0.6pt] (JJ2_R.north) -- (7.1, 2.7) -- (6.5, 2.7);
    
    % Upward connection
    \node[circ] at (5, 3.5) {};
    \draw[line width=0.6pt] (5, 3.5) -- (5, 4.5) coordinate (C2_top);
    
\end{scope}

% ---------------------------------------------------
% COUPLING CAPACITORS 
% ---------------------------------------------------
% C1 (1 to Coupler)
\draw[line width=0.6pt] (C1_top) -- (-2.5, 4.5) to[C, l=$C_{1}$] (-0.5, 4.5) -- (Cc_top);

% C2 (2 to Coupler)
\draw[line width=0.6pt] (Cc_top) -- (0.5, 4.5) to[C, l=$C_{2}$] (2.5, 4.5) -- (C2_top);

% C12 (Two-qubit direct coupling)
\draw[line width=0.6pt] (C1_top) -- (-5, 6.5) -- (-1.5, 6.5) to[C, l=$C_{12}$] (1.5, 6.5) -- (5, 6.5) -- (C2_top);

\end{tikzpicture}
  
  \caption{Circuit consisting of qubits 1 (red, left) and 2 (blue, right) coupled directly between them through capacitor $C_{12}$, indirectly through the tunable coupler (black, center) through capacitors $C_1$ and $C_2$, and to individual and independent external control lines (magenta), through capacitors $C_{\mathrm{d},1}$ and $C_{\mathrm{d},2}$.}
  \label{fig:two_qubit_circuit}
\end{figure}

Combining Eqs.~(\ref{eq:H0}),~(\ref{eq:Hint}), and~(\ref{eq:HAC}), we obtain the full Hamiltonian of the system to be
\begin{align}\label{eq:Htotal}
    H(t) = H_0 + H_\mathrm{int} + H_\mathrm{d}(t).
\end{align}
 The entire system is depicted in Fig.~\ref{fig:two_qubit_circuit}. For more details on the dependency relations of the coupling constants with the capacitance values we refer to ~\cite{marxer2023IQM}.
 
 Besides the interactions within the circuit, each qubit is also coupled to its surrounding electromagnetic environment, leading to thermal relaxation and excitation processes \cite{babu2023quantum}. Such process can be described by the quantum channel~\cite{petruccione}
 \begin{align}
     \mathcal{D}_k(\rho) = \gamma(1+\bar{n}) D[a_k](\rho) + \gamma \bar{n} D[a_k^\dagger](\rho)
 \end{align}
 acting on each qubit subspace, where $\gamma$ is the relaxation rate, $\bar{n}$ is the average photon number in the cavity, and $D[a_k]$ is given by
 \begin{align}
     D[a_k](\rho) \equiv a_k \rho a_k^\dagger - \frac{1}{2}\acomm{a_k^\dagger a_k}{\rho},
 \end{align}
where $\{\cdot,\cdot\}$ denotes the anti-commutator. The evolution of the  system will then be described by the Lindblad equation
 \begin{align}\label{eq:Liouville_equation}
     \frac{\mathrm{d}\rho}{\mathrm{d}t} = -i \comm{H(t)}{\rho(t)} + \sum_{k=1}^2 \mathcal{D}_k(\rho(t)),
 \end{align}
 where $[\cdot,\cdot]$ denotes the commutator.

\subsection{Effective Hamiltonian}\label{subsec:effective_Hamiltonian}

Equation~(\ref{eq:Liouville_equation}) is not restricted to a finite dimension since the qubits are still being considered as an infinite-dimensional Duffing oscillator. The first step is therefore to truncate the qubits subspace. For the purposes of this work, we will consider that each qubit, and the coupler, has a single-dimensional leakage space, that is, each qubit subspace is spanned by $\{\ket{0}, \ket{1}, \ket{2}\}$. Thus, considering the energy of the vacuum state to be zero, the native Hamiltonian $H_0$, Eq.~(\ref{eq:H0}), can be explicitly written as
\begin{equation}\label{eq:H0_truncated}
\begin{aligned}
    H_0 &= \omega_c \ketbra{1_c}{1_c} + (2\omega_c + \alpha_c) \ketbra{2_c}{2_c} \\ &\quad + \sum_{k=1}^2 \omega_k \ketbra{1_k}{1_k} + (2\omega_k + \alpha_k) \ketbra{2_k}{2_k}.
\end{aligned}
\end{equation}
Having a system of three 3-level subsystems coupled means that the unitary part of the evolution is described by elements of the $\mathrm{SU}(3^3) = \mathrm{SU}(27)$ Lie group, which has an algebra with dimension $728$.~\cite{helgason1979} Performing a geodesic search in such a high-dimensional space is unfeasible, so further considerations must be made.

In most quantum computing applications, we are generally interested in accurately controlling the state of the two qubits, while using the coupler exclusively as a tool. For that reason, we consider the coupler to be far detuned from the natural frequency of the qubits, that is, $\omega_c \gg \omega_k$, for $k=1,2$. This means that the coupler is only virtually populated, effectively remaining in the ground state during the evolution, and only inducing perturbative corrections to the qubits' frequencies and the coupling constants. This approximation is achieved by means of a Schrieffer-Wolff (SW) transformation.~\cite{bravyi2011schrieffer} One last approximation is done by considering that the two qubits are closely resonant, that is, $\omega_1 \approx \omega_2$, which enables us to perform a rotating-wave approximation (RWA), eliminating terms that oscillate too fast when the system is transformed to the co-rotating frame with the frequency of the qubits. 

After the SW, we effectively have a Hamiltonian involving only the degrees of freedom of the two qubits, with perturbatively corrected frequencies and anharmonicities, given by
\begin{align}
    \widetilde{\omega}_k &\equiv \omega_k + \frac{g_k^2}{\Delta_k} + \frac{g_k^2}{\Sigma_k} - \frac{2g_k^2}{\Sigma_k + \alpha_k}, \label{eq:omegaktilde} \\
    \widetilde{\alpha}_k &\equiv \alpha_k - \frac{2g_k^2}{\Delta_k} + \frac{2g_k^2}{\Delta_k+\alpha_k} - \frac{g_k^2}{\Sigma_k} + \frac{4g_k^2}{\Sigma_k+\alpha_k},\label{eq:alphaktilde}
\end{align}
with $\Delta_k \equiv \omega_k - \omega_c$, and $\Sigma_k \equiv \omega_k + \omega_c$. We will refer to these quantities as dressed frequencies and dressed anharmonicities respectively.
The truncation of the system up to the second excited state allows us to rewrite the drive Hamiltonian of Eq.~(\ref{eq:HAC}) as
\begin{align}\label{eq:Hd_original}
    H_\mathrm{d}(t) &= \sum_{k=1}^2 -i c_k V_{\mathrm{d},k}(t) \nonumber \\ &\quad \times \mleft[ \ketbra{0_k}{1_k} - \ketbra{1_k}{0_k} + \sqrt{2} \left( \ketbra{1_k}{2_k} - \ketbra{2_k}{1_k} \right) \mright].
\end{align}
Using Eq.~(\ref{eq:Vdk_original}), it is direct to check that transforming the system to the co-rotating frame using the operator $U_\mathrm{ho} \equiv \exp(-i H_\mathrm{ho} t)$, where
$H_\mathrm{ho} \equiv \sum_k \omega_k \ketbra{1_k}{1_k} + 2 \omega_k \ketbra{2_k}{2_k}$, and defining the matrices
\begin{align}\label{eq:lambx_and_lamby}
    \lambda_x \equiv \frac{1}{\sqrt{3}} \begin{pmatrix}
        0 & 1 & 0 \\ 1 & 0 & \sqrt{2} \\ 0 & \sqrt{2} & 0
    \end{pmatrix}, \;
    \lambda_y \equiv \frac{1}{\sqrt{3}} \begin{pmatrix}
        0 & -i & 0 \\ i & 0 & -i\sqrt{2} \\ 0 & i\sqrt{2} & 0
    \end{pmatrix},
\end{align}
will yield a control Hamiltonian given by
\begin{align}\label{eq:Hd}
    H_\mathrm{c}(t) = \sum_{k=1}^2 \left[\Omega_{x,k}(t) \lambda_{x,k} + \Omega_{y,k}(t) \lambda_{y,k} \right].
\end{align}
where we have defined the Rabi frequencies
\begin{align}\label{eq:Rabi}
    \Omega_{x,k}(t) &\equiv \tfrac{\sqrt{3}}{2} c_k \left[ V_{I,k}(t) \sin{\theta_k(t)} - V_{Q,k}(t) \cos{\theta_k(t)} \right], \nonumber \\
    \Omega_{y,k}(t) &\equiv \tfrac{\sqrt{3}}{2} c_k \left[ V_{I,k}(t) \cos{\theta_k(t)} + V_{Q,k}(t) \sin{\theta_k(t)} \right],
\end{align}
with $\theta_k(t) \equiv (\omega_{\mathrm{d},k} - \widetilde{\omega}_k )t + \phi_k$. The factor of $1/\sqrt{3}$ in the definitions of $\lambda_x$ and $\lambda_y$ are just a convenient choice such that these matrices are normalized to $\tr(\lambda_x^2) = \tr(\lambda_y^2) = 2$, which is also satisfied by the Gell-Mann matrices, that form a basis for $\mathfrak{su}(3)$.

Combining Eq.~(\ref{eq:Hd}) with the post SW transformation of the interaction Hamiltonian, and changing to the interaction picture, yields the Hamiltonian
\begin{widetext}
\begin{equation}\label{eq:Heff}
\begin{aligned}
    H_\mathrm{eff}(t) &= \sum_{k=1}^2 \widetilde{\alpha}_k \ketbra{2_k}{2_k} + \sum_{k=1}^2 \left[ \Omega_{x,k}(t) \lambda_{x,k} + \Omega_{y,k}(t) \lambda_{y,k}, \right] \\
    &\quad + e^{-i \widetilde{\Delta} t} \left[ \Gamma_{\downarrow\downarrow} \ketbra{10}{01} + \sqrt{2} \left( \Gamma_{\downarrow\uparrow} \ketbra{11}{02} + \Gamma_{\uparrow\downarrow} \ketbra{20}{11} \right) + 2 \Gamma_{\uparrow \uparrow} \ketbra{21}{12} \right] + \mathrm{h.c.},
\end{aligned}
\end{equation}
\end{widetext}
The entry $\ket{2_k}$ denotes the $\ket{2}$ state for the $k$-th qubit, and $\ket{\ell m}$ should be interpreted as $\ket{\ell_1} \otimes \ket{m_2}$. A summary of the quantities in Eq.~(\ref{eq:Heff}) follows. $\Gamma_{jk}$ with $j,k \in \{\uparrow, \downarrow\}$, are the effective perturbed coupling constants. It can be shown that they are given by
\begin{equation*}
    \begin{aligned}
    \Gamma_{\downarrow \downarrow} &\equiv g_{12} + \tfrac{g_1g_2}{2} \left( \tfrac{1}{\Delta_1} + \tfrac{1}{\Delta_2} - \tfrac{1}{\Sigma_1} - \tfrac{1}{\Sigma_2} \right), \\
    \Gamma_{\downarrow \uparrow} &\equiv g_{12} + \tfrac{g_1g_2}{2} \left( \tfrac{1}{\Delta_1} + \tfrac{1}{\Delta_2+\alpha_2} - \tfrac{1}{\Sigma_1} - \tfrac{1}{\Sigma_2+\alpha_2} \right), \\
    \Gamma_{\uparrow \downarrow} &\equiv g_{12} + \tfrac{g_1g_2}{2} \left( \tfrac{1}{\Delta_1+\alpha_1} + \tfrac{1}{\Delta_2} - \tfrac{1}{\Sigma_1+\alpha_1} - \tfrac{1}{\Sigma_2} \right), \\
    \Gamma_{\uparrow \uparrow} &\equiv g_{12} + \tfrac{g_1g_2}{2} \left( \tfrac{1}{\Delta_1+\alpha_1} + \tfrac{1}{\Delta_2+\alpha_2} - \tfrac{1}{\Sigma_1+\alpha_1} - \tfrac{1}{\Sigma_2+\alpha_2} \right),
    \end{aligned}
\end{equation*}
and $\widetilde{\Delta} \equiv \widetilde{\omega}_2 - \widetilde{\omega}_1$ is the detuning between the qubits using the dressed frequencies.
For a detailed derivation of all results from Eq.~(\ref{eq:omegaktilde}) to~(\ref{eq:Heff}), see Appendix~\ref{appendix:calculations}. The arrow symbols $\downarrow$ and $\uparrow$ here are being used to heuristically refer to the ``lower'' transition (between levels $\ket{0}$ and $\ket{1}$), and ``higher'' transition (between levels $\ket{1}$ and $\ket{2}$), respectively. The numerical values chosen for our purposes were $\omega_1/(2\pi) = \omega_2/(2\pi) = 4.854$ GHz, $\alpha_1/(2\pi) = \alpha_2/(2\pi) = - 191$ MHz, $g_1/(2\pi) = 69.6$ MHz, $g_2/(2\pi) = 62.2$ MHz, and $g_{12}/(2\pi) = 1.7$ MHz. This choice of parameters was made based on the reported values for one of IQM's quantum computers ~\cite{marxer2026above}.
\begin{figure}[]
    \centering
    \includegraphics[scale=0.6]{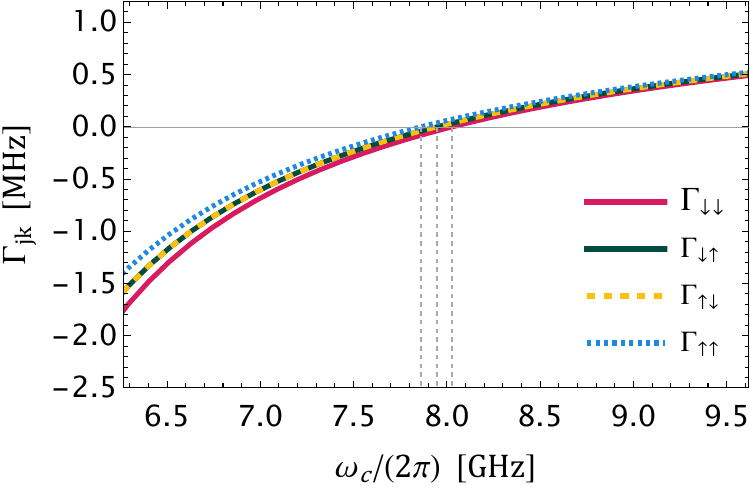}
    \caption{Effective coupling constants of Eq.~(\ref{eq:Heff}) as a function of the frequency of the tunable coupler. The gray dashed vertical lines indicate the values for which each coupling is null. The optimal value for minimal interaction is $\omega_c/(2\pi) = 7.945$ GHz, for which we get $\abs{\Gamma_{\downarrow\downarrow}}$ and $\abs{\Gamma_{\uparrow\uparrow}}$ in the order of $240$ Hz, and $\abs{\Gamma_{\downarrow\uparrow}} = \abs{\Gamma_{\uparrow\downarrow}} \approx 4$ Hz. For the time scale of single-qubit gates, these values correspond to the two qubits being effectively decoupled. For the $i$SWAP interaction, we select $\omega_c/(2\pi) = 5.862$ GHz, which is still comfortably within the dispersive regime while allowing for exchange interactions.}
    \label{fig:effective_coupling_constants}
\end{figure}
Figure~\ref{fig:effective_coupling_constants} shows the effective coupling constants as a function of the coupler frequency. Notice that it is possible to select a value for $\omega_c$ such that the two qubits become effectively decoupled. The optimal value for the coupler frequency was set by averaging the value of $\omega_c/(2\pi)$ that causes each coupling to be turned off completely. The found value was $7.945$ GHz. For this value, we get $\abs{\Gamma_{\downarrow\downarrow}}$ and $\abs{\Gamma_{\uparrow\uparrow}}$ around $240$ Hz, while $\abs{\Gamma_{\downarrow\uparrow}}$ and $\abs{\Gamma_{\uparrow\downarrow}}$ are around $4$ Hz. As it will be shown in the next section, single-qubit operations are set to occur in tens of nanoseconds. At this scale, these coupling constant values can be safely truncated to zero, meaning the qubits are effectively decoupled. For an $i$SWAP interaction to happen, we set $\omega_c/(2\pi) = 5.862$ GHz. This value corresponds to a detuning of over $1$ GHz between the coupler and the qubits, meaning we are in the dispersive regime, while allowing an exchange interaction to happen between the qubits.

\subsection{Algebra and control subspace}\label{subsec:2qt-algebra}

Here we  shortly describe how this physical model is represented within the geometrical framework shown in section~\ref{sec:optimal_control}. Since the coupler is considered to be in the dispersive regime, influencing only perturbatively the qubit frequencies, we effectively have a system composed by two 3-level systems, which has $\mathrm{SU}(3^2) = \mathrm{SU}(9)$ as unitary group. The natural choice of basis for the algebra $\mathfrak{su}(9)$ is composed of strings of Gell-Mann matrices with weights 1 and 2, that is, 
\begin{align}
    \{\lambda_j \otimes \mathbb{I}, \mathbb{I} \otimes \lambda_j, \lambda_j \otimes \lambda_k\},
\end{align}
where $j$ and $k$ vary from 1 to 8, corresponding to the eight Gell-Mann matrices, and $\mathbb{I}$ is the single-qutrit identity operator. We have that $\dim[\mathfrak{su}(9)] = 80$, and thus, the costate $\Lambda(0)$ consists of $80$ real numbers. The control subspace is spanned by $\lambda_x = \tfrac{1}{\sqrt{3}}(\lambda_1 + \sqrt{2}\lambda_6)$ and $\lambda_y = \tfrac{1}{\sqrt{3}}(\lambda_2 + \sqrt{2}\lambda_7)$, defined in Eq.~(\ref{eq:lambx_and_lamby}), acting on each qubit separately. Therefore we adopt
\begin{align}\label{eq:distribution_basis}
    \{\lambda_x \otimes \mathbb{I}, \lambda_y \otimes \mathbb{I}, \mathbb{I} \otimes \lambda_x, \mathbb{I} \otimes \lambda_y\}
\end{align}
to be the operators that span the distribution $\mathcal{V}$, and thus a basis for this vector space.

Using the same terms introduced in section~\ref{sec:optimal_control}, we have a physical system with Lie algebra $\mathfrak{g} = \mathfrak{su}(9)$, such that $\mathrm{dim}(\mathfrak{g}) = 80$, and control subspace $\mathcal{V}$ spanned by Eq.~(\ref{eq:distribution_basis}), meaning $\mathrm{dim}(\mathcal{V}) = 4$.

\subsection{Leakage, seepage and average fidelity}\label{subsec:leakage_and_seepage}

The analysis of leakage and seepage is done following ~\cite{Wood2024}. First we define the projectors onto the computational, and leakage subspaces, respectively by
\begin{align}
    \mathrm{P}_\mathrm{c} \equiv \sum_{j,k=0}^1 \ketbra{jk}{jk}, \quad
    \mathrm{P}_\ell \equiv \mathbf{1} - \mathrm{P}_\mathrm{c},
\end{align}
where $\mathbf{1}$ denotes the identity operator for the entire two-qudit space. The leakage rate that some quantum channel $\mathcal{E}$ induces in some initial state $\rho$ is then defined as the amount of population in the leakage subspace after the action of $\mathcal{E}$. Operationally, it is written as
\begin{align}\label{eq:L1}
    L_1(\mathcal{E}(\rho)) \equiv \tr[ \mathrm{P}_\ell \mathcal{E}(\rho) ] = 1 - \tr[\mathrm{P}_\mathrm{c} \mathcal{E}(\rho)].
\end{align}
Conversely, the seepage rate of a channel is defined as the amount of population in the computational subspace after the action of $\mathcal{E}$
\begin{align}\label{eq:L2}
    L_2(\mathcal{E}(\rho)) \equiv \tr[\mathrm{P}_\mathrm{c} \mathcal{E}(\rho)].
\end{align}
Notice that in order for $L_1$ and $L_2$ to account for leakage and seepage exclusively due to the action of a quantum channel, it is implied that $\rho$ in Eq.~(\ref{eq:L1}) has support in the computational subspace, and in Eq.~(\ref{eq:L2}), it has support in the leakage subspace.

% Notice that, although the geodesic search restricts the optimization to the computational subspace plus a single-dimensional leakage level for each qubit, the leakage analysis and seepage analysis can be made considering an arbitrary dimension for the leakage subspace. The results would then be interpreted as how efficient is the optimal control tailored to a single-dimensional leakage level space when used in a context where higher energy levels can be possibly populated.

To have an analysis independent of state preparation, we consider the leakage and seepage rates averaged over the Haar measure of all possible initial states. Using $d_\ell$ to denote the dimension of the leakage subspace we have
\begin{align}
    \overline{L}_1(\mathcal{E}) &= \int \mathrm{d} \psi_\mathrm{c} \, L_1(\mathcal{E}(\ketbra{\psi_\mathrm{c}}{\psi_\mathrm{c}})) = L_1 \mleft( \mathcal{E}\mleft( \frac{\mathrm{P}_\mathrm{c}}{d_\mathrm{c}} \mright) \mright), \label{eq:avg_leakage} \\ 
    \overline{L}_2(\mathcal{E}) &= \int \mathrm{d} \psi_\ell \, L_2(\mathcal{E}(\ketbra{\psi_\ell}{\psi_\ell})) = L_2 \mleft( \mathcal{E}\mleft( \frac{\mathrm{P}_\ell}{d_\ell} \mright) \mright),  \label{eq:avg_seepage}
\end{align}
where $\ketbra{\psi_\mathrm{c}}{\psi_\mathrm{c}}$ and $\ketbra{\psi_\ell}{\psi_\ell}$ are pure states in the computational and leakage subspaces, respectively.

Besides the average leakage and seepage rates, we also analyze the average fidelity, which can be calculated as~\cite{nielsen2002simple, bowdrey2002fidelity}
\begin{align}\label{eq:avg_fidelity}
    \overline{F} \equiv \frac{d_\mathrm{c} F_\mathrm{pro}(\mathcal{E}) + 1 - \overline{L}_1(\mathcal{E})}{d_\mathrm{c} + 1}.
\end{align}
$F_\mathrm{pro}(\mathcal{E})$ is the process fidelity, given by
\begin{align}
    F_\mathrm{pro}(\mathcal{E}) \equiv \frac{1}{d_\mathrm{c}^2} \abs{\tr[\mathcal{U}^\dagger \mathcal{X}_\mathcal{E}]}^2,
\end{align}
where $\mathcal{U}$ is the ideal, noiseless, $\mathrm{SU}(4)$ target unitary, in superoperator form, and $\mathcal{X}_\mathcal{E}$ is the superoperator form of the quantum channel acting on the computational subspace of the two-qubit system. It can be obtained by rewriting Eq.~(\ref{eq:Liouville_equation}) in superoperator form, and then extracting the submatrix corresponding exclusively to the computational subspace from the superoperator. More specifically, solving Eq.~(\ref{eq:Liouville_equation}) yields a solution that can be written in the form
\begin{align*}
    \rho(t) = \mathcal{X}(t) \rho(0)
\end{align*}
for some superoperator $\mathcal{X}(t)$. Using $\tau$ to denote the evolution time, $\mathcal{X}_\mathcal{E}$ is obtained by taking the submatrix of $\mathcal{X}(\tau)$ that acts on the computational subspace of the two qubits.

In the context of this work, we are always assuming $d_\mathrm{c} = 4$, meaning the computational space is always spanned by $\{\ket{00}, \ket{01}, \ket{10}, \ket{11}\}$. However, although the process of finding the optimal $H_\mathrm{c}(t)$ considers only the states spanned by the first leakage level $\ket{2}$, that is, the space spanned by $\{\ket{02}, \ket{12}, \ket{20}, \ket{21}, \ket{22}\}$ meaning $d_\ell = 5$, the analysis of the quantities defined in this section need not be restricted to these dimensions. The ladder operators can be truncated to any level $\ket{n}$ once the functions $\Omega_{x,k}(t)$ and $\Omega_{y,k}(t)$ are known for $k = 1,2$. This was in fact tested by including the levels $\ket{3}$ and $\ket{4}$ in the dynamics, however, addition of $\ket{3}$ introduced only negligible differences, and for $\ket{4}$, any differences were imperceptible. 

\section{Results}\label{sec:results}

\subsection{Single-qubit $R_x(\pi/2)$ gate}

\begin{figure}[]
    \centering
    \includegraphics[scale=0.6]{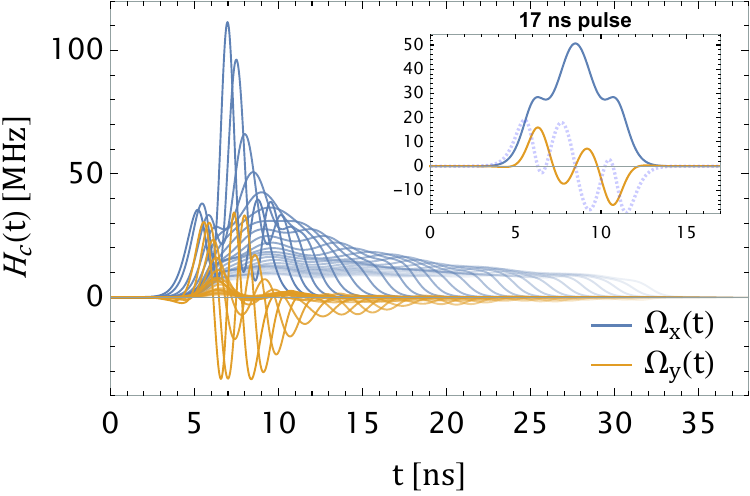}
    \caption{Single-qubit pulses resulting in the $R_x(\pi/2)$ rotation operation with gate time $\tau$ for $24$ different values ranging from $\tau = 14$ ns to $\tau = 38$ ns. The curves are shown with decreasing opacity as the pulse duration increases. Inset we highlight the specific $17$ ns pulse. In dashed blue we show $-(1/\alpha) \mathrm{d}\Omega_x(t)/\mathrm{d}t$, showing that the geodesic optimal pulse does not result in $\Omega_y(t)$ proportional to the first derivative of $\Omega_x(t)$, differently from DRAG pulses.}
    \label{fig:X90_pulses}
\end{figure}

\begin{figure}[]
    \centering
    \includegraphics[scale=0.6]{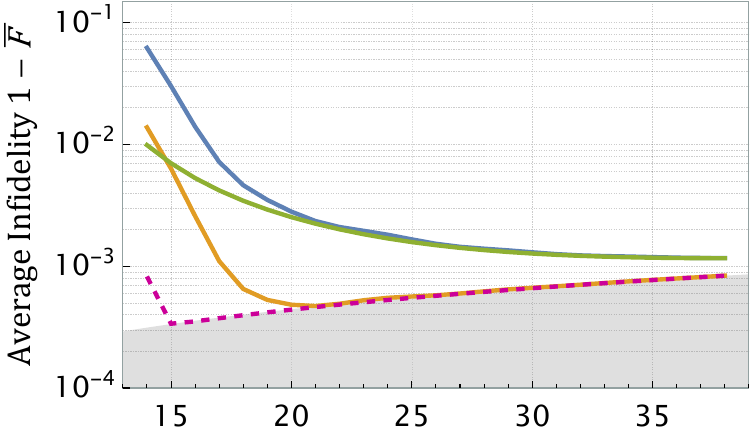}
    \includegraphics[scale=0.6]{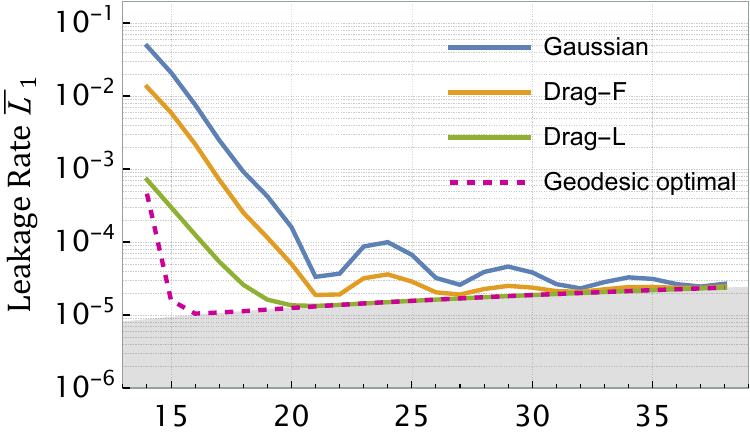}
    \includegraphics[scale=0.6]{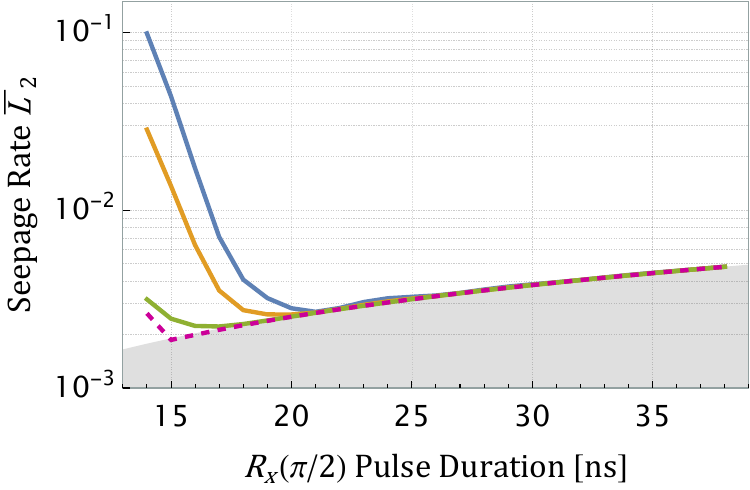}
    \caption{Average infidelity, leakage, and seepage rates as functions of the duration of a single-qubit $R_x(\pi/2)$ pulse using different pulse types. The shaded grey area corresponds to the limit set by the simulated thermal dissipation process. The optimized pulses through the geodesic search method show values for the three analyzed quantities minimized to the limit set by thermal dissipation even for short pulses.}
    \label{fig:avg-fid-single-qb}
\end{figure}

The first set of results we present explore how the pulses obtained from this optimal control method compares to other, well known pulses. We use a single-qubit $\pi/2$ rotation around the Bloch sphere X axis, given by
\begin{align}
    R_x(\pi/2) = \frac{1}{\sqrt{2}}\begin{pmatrix}
    1 & -i \\ -i & 1 
    \end{pmatrix}
\end{align}
for this comparison. The other pulses analyzed are a simple Gaussian pulse, as well as DRAG. While for the former we have $\Omega_x(t)$ given by a Gaussian shaped pulse with $\Omega_y(t) = 0$, DRAG pulses consists of adding a correction proportional to the derivative of $\Omega_x(t)$, of the form
\begin{align}
    \Omega_y(t) = -\frac{\delta}{\alpha} \frac{\mathrm{d}}{\mathrm{d}t} \Omega_x(t),
\end{align}
where $\alpha$ is the qubit's anharmonicity, and $\delta$ is called the DRAG parameter, which can vary from $0$ to $1$. Here we consider two different values: $\delta=0.5$, which optimizes gate fidelity (DRAG-F), and $\delta = 1$, which minimizes leakage and seepage (DRAG-L).~\cite{chen2016measuring} Notice that the Gaussian pulse is equivalent to a DRAG pulse with $\delta = 0$.

In order to be consistent with the geodesic optimal pulses, we set the Gaussian and DRAG pulses to be preceded and followed by an idling time of $4$ ns, such that the Gaussian rise and fall starts and ends roughly at the same time as the geodesic optimal pulses using $r=2.4$, and $t_\mathrm{off} = 6$ ns in the envelope function defined in Eq.~(\ref{eq:envelope}). For consistency, we include the idling time preceding and following the Gaussian and DRAG pulses in the total pulse duration. We solve Eq.~(\ref{eq:geodesiceq_this_work}) for pulse durations ranging from $14$ ns to $38$ ns in steps of $1$ ns, totalizing $24$ different pulses for the $R_x(\pi/2)$ gate. This means that for the Gaussian and DRAG pulses, the actual Gaussian envelope ranges from $6$ ns to $30$ ns. Without loss of generality, we apply the single-qubit operations to the first qubit, while the second one is kept idle. For that we tune the coupler to $\omega_c = 7.945$ GHz, which effectively decouples the two qubits. The optimization process is then simplified to a search in $\mathrm{SU}(3)$ instead of $\mathrm{SU}(9)$.

Figure~\ref{fig:X90_pulses} shows all the control functions numerically obtained with the geodesic method.
%with pulse duration ranging from $14$ ns to $38$ ns.
%numerically \tp{obtained} control functions $\Omega_x(t)$ and $\Omega_y(t)$ with the geodesic method.
Inset we show specifically the pulse with duration $17$ ns. In dashed line it is shown the function $-(1/\alpha) \mathrm{d}\Omega_x/\mathrm{d}t$, showing that the $\Omega_y(t)$ is not proportional to the first derivative, differently from DRAG pulses. It is a possibility, however, that $\Omega_y(t)$ can be approximated by considering higher order derivatives of $\Omega_x(t)$, and this is an attractive subject for future research following this work. In case it is possible to approximate $\Omega_y(t)$ using higher order derivative corrections, this would be an interesting result connecting sub-Riemannian geodesic search problem to the design of generalized DRAG pulses.

In Fig.~\ref{fig:avg-fid-single-qb}, we compare the performance of the geodesic optimal pulses with the Gaussian and DRAG pulses, analyzing the values for average gate infidelity, calculated as $1 - \overline{F}$, with $\overline{F}$ given by Eq.~(\ref{eq:avg_fidelity}), and average leakage and seepage rates, given by Eqs.~(\ref{eq:avg_leakage}) and~(\ref{eq:avg_seepage}). The shaded grey regions indicate the limits set by the thermal dissipation process, which are achieved by setting relaxation factor $\gamma/(2\pi) = 10$ kHz and average photon number $\overline{n} = 0.01$, similar to Ref.~\cite{Wood2024}. Notice that for all three analyzed quantities, the geodesic optimal pulse performs better for fast pulses, that is, less than $20$ ns, with DRAG performance matching only for longer pulses. The results show that the geodesic optimal pulses can be used to combine the effects of both DRAG-F and DRAG-L, and to saturate the values of average infidelity, leakage and seepage to the thermal limit for a pulse as short as $17$ ns. The obtained values for this pulse duration were $\overline{F} = 0.9996$, $\overline{L}_1 = 1.08 \times 10^{-5}$, and $\overline{L}_2 = 2.12 \times 10^{-3}$.

These results considered a physical system consistent with current superconducting platforms, in which it is possible to select the frequency of the couplers, and in some cases, even adjust the frequencies of the qubits themselves by the same means. The only possible obstacles in practical implementations of the method would thus be the level of precision one can synthesize pulses similar to the ones shown in Fig.~\ref{fig:X90_pulses}, using arbitrary wave generators (AWG), as well as the level of precision in measurements of the numerical values for the coupling terms in the Hamiltonian.

\subsection{Two-qubit $i$SWAP gate}

\begin{figure}[]
    \centering
    \includegraphics[scale=0.6]{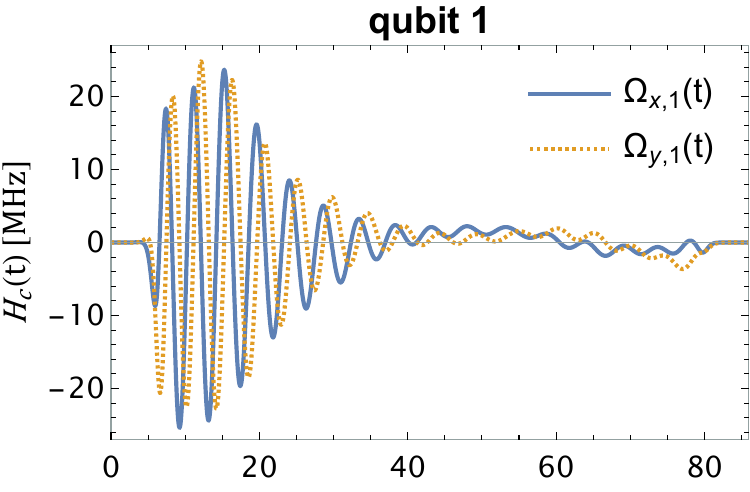}
    \includegraphics[scale=0.6]{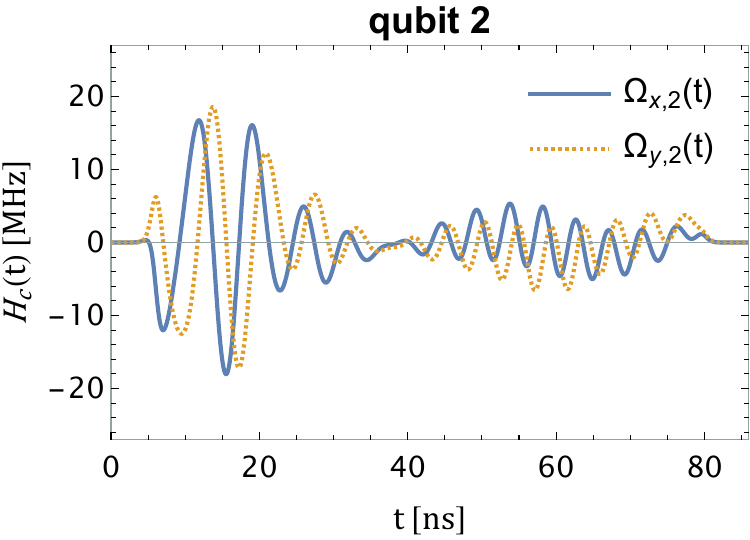}
    \caption{Single-qubit pulses applied on each qubit for the synthesis of an $i$SWAP two-qubit gate. Since the optimization is done in the frame co-rotating with the two-qubits' frequencies, the oscilations indicate the applied pulses must be off-resonant with the qubits.}
    \label{fig:2qb-iSWAP-pulses}
\end{figure}

%When the two qubits are in a resonant regime, with the coupler frequency set to a value for which exchange interactions are possible albeit still being in the dispersive regime with the qubits, it is possible to achieve an $i$SWAP interaction without using any external control. \ahs{This kind of interaction was recently shown to be particularly important for quantum low-density parity check (qLDPC) codes in the context of QEC in superconducting qubits~\cite{gu2026nearest}}.

When the two qubits are in a resonant regime and the coupler frequency is tuned such that exchange interactions become possible, while still maintaining dispersive coupling to the qubits, an $i$SWAP interaction can be generated naturally without external control fields. Such interactions have recently been shown to play an important role in the implementation of quantum low-density parity-check codes (qLDPC) for quantum error correction in superconducting-qubit platforms~\cite{gu2026nearest}.

The selected coupler frequency value for this purpose was $\omega_c = 5.862$ GHz, which corresponds to a detuning of $\sim 1$ GHz from the qubits while causing effective couplings $\Gamma_{jk}$ in the order of tens of MHz. With these parameters, an $i$SWAP operation can be performed in $79.5$ ns with average fidelity $\overline{F} \approx 0.95$.

Given the promising performance of the geodesic optimal pulses for maximizing average fidelity and minimizing leakage and seepage for single qubit gates, it is a pertinent analysis to check whether the same process can be used to either increase the fidelity for $i$SWAP or decrease the required gate time. By testing gate times around $80$ ns, the best performance was encountered when the gate time was set to $86$ ns, for which an $i$SWAP gate with $\overline{F} = 0.9942$ was achieved, requiring only an increasing of $6$ ns to the required evolution time. By turning on the external control fields, the leakage rate actually increases slightly compared to the control-free evolution, while the seepage rate was kept effectively the same. This is not necessarily a problem since the increase in leakage is negligible compared to the increase in the average fidelity. For comparisons, in the control-free evolution, the rates were $\overline{L}_1 = 4.55 \times 10^{-4}$ and $\overline{L}_2 = 8.98 \times 10^{-3}$, while with the external control they were $\overline{L}_1 = 6.69 \times 10^{-4}$ and $\overline{L}_2 = 8.94 \times 10^{-3}$. These are negligible differences compared to the over $0.04$ increase in average fidelity.

In Fig.~\ref{fig:2qb-iSWAP-pulses} we show the geodesic optimal single-qubit pulses applied on each qubit for the $i$SWAP evolution. One noticeable major difference from the pulses shown in Fig.~\ref{fig:X90_pulses} is the presence of fast oscillations between positive and negative values, while in the pulses for the $R_x(\pi/2)$ gate they consisted mostly of rise-and-fall generally. The explanation is that, since all simulations are performed in the frame co-rotating with the qubits' dressed frequencies, fast oscilations in $\Omega_x(t)$ and $\Omega_y(t)$ just means that such pulses are off-resonant, which is expected, since the control in this case would mostly be used to perform real-time corrections to the frequency of the qubits, instead of performing any single-qubit rotations. Since $g_1$ and $g_2$ are considered to slightly mismatch, the dressed frequencies of Eq.~(\ref{eq:omegaktilde}) will be such that the qubits will be out of perfect resonance.

\begin{figure}[]
    \centering
    \includegraphics[scale=0.6]{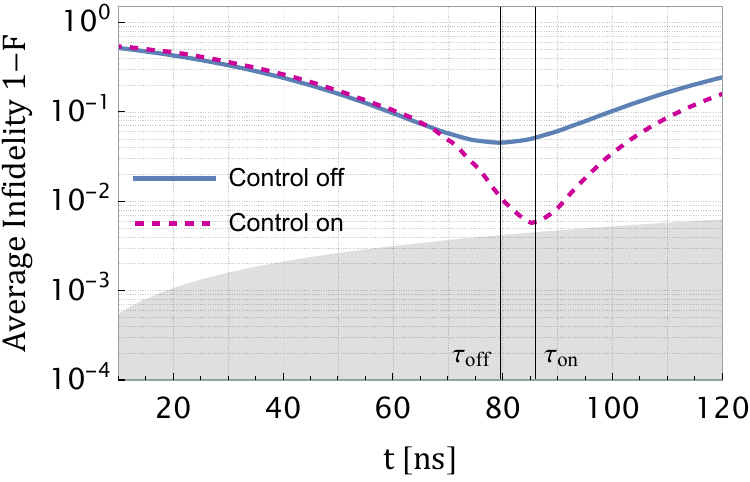}
    \caption{Average infidelity for the $i$SWAP operation as function of evolution time. In complete absence of any signal from the control line, an $i$SWAP with $\overline{F} = 0.9548$ is achieved at $t = 79.5$ ns. By turning on the control line with the geodesic optimized pulses, it is possible to synthesize the $i$SWAP with $\overline{F} = 0.9942$ at $t = 86$ ns. The shaded grey area corresponds to the limit set by the simulated thermal dissipation process. The two vertical lines only indicate precisely the values in time $\tau_\mathrm{off} = 79.5$ ns and $\tau_\mathrm{on} = 86$ ns for which the infidelity is minimum.}
    \label{fig:iSWAP-avf-inf}
\end{figure}

In Fig.~\ref{fig:iSWAP-avf-inf}, we compare the time evolutions with and without the optimal control fields. As already mentioned, the external control in this case is used to cause real-time corrections in the qubit frequencies such that the swap interaction can be performed with higher fidelity, albeit taking slightly more time to complete. These results show that this optimal control method can be used both for single-qubit rotation operations, as well as dynamically fine tuning qubit interaction parameters in real time during evolution. Evidently, in platforms where the qubits' frequencies can be controlled at any point in time, this time-dependent control could be performed using other methods, involving time-dependent magnetic flux variations for example. Exploring the use of this method to obtain optimal protocols for magnetic flux instead of external micro-wave range pulses is an interesting possible path for future research.

% The result showing how the optimal control can be used to improve average fidelity for the $i$SWAP gate go here. The control actually introduces slightly more leakage to the system than evolving with the interaction Hamiltonian alone, but compensates by increasing the average gate fidelity considerably. Similar to what DRAG-F pulse is supposed to do.

\section{Conclusions}\label{sec:conclusion}

In this work we have adapted the optimal control method based on sub-Riemannian geodesic search, introduced in ~\cite{morazotti2024optimized, da2025monte}, for a realistic superconducting two-qubit system capacitively coupled directly and indirectly via a tunable coupler for suppressing leakage errors. More specifically, we enforced the constraint that the pulses should start and end at zero directly in the sub-Riemannian geodesic equation. This ensures smooth turn-on and turn-off transitions, avoiding the broadening of the control function in the frequency domain, and thus, minimizing spurious excitations to non-computational states. We remark that energy minimization is the only condition considered for the control pulses, with maximization of fidelity only enforced as a search of the best solution among all possible energy optimal paths. Furthermore, because the search space encompasses both the computational and leakage subspaces, solutions that maximize process fidelity also inherently minimize leakage.

For simulating leakage processes we considered two main sources for the noise: coherent excitations caused by the control field itself, and incoherent thermal dissipation.
We started considering the entire two-qubit system with the tunable coupler, then derived an effective interaction Hamiltonian by setting the coupler to interact in the dispersive regime. This allowed us to consider optimization restricted to unitary evolution operators belonging to the $\mathrm{SU}(9)$ group.

For evaluating single-qubit operations performance, we chose the $R_x(\pi/2)$ rotation operation. We showed that the geodesic optimal pulses outperform DRAG pulses, especially in the regime of very short pulses. For instance, we showed a $17$ ns pulse with smooth rise and fall curves that enables saturation of the levels of average gate infidelity, leakage, and seepage rates to the limit set by the thermal dissipation process. We showed that the method is able to combine the benefits of both DRAG-F and DRAG-L pulses, meaning it can simultaneously optimize for gate fidelity and for minimizing leakage errors.

Extending the approach to two-qubit operations, we used the $i$SWAP operation, which is naturally compatible with the derived effective Hamiltonian in the sense that in the absence of any control, the $i$SWAP operation can be executed with an average fidelity of around $0.95$ in $79.5$ ns. We showed that it is possible to use the geodesic optimal pulses for dynamically applying real-time corrections to the qubits' frequencies, causing the swap exchange to have a higher fidelity albeit requiring a slightly longer gate time. 

In summary, this work shows that an approach based on sub-Riemannian geodesic search can be used to perform energy optimal control over realistic superconducting qubit platforms that simultaneously deliver high fidelity quantum operations with low leakage levels, serving as a possible improvement from quantum control based on DRAG pulse applications. An interesting open question is whether there are any connections between the single-qubit geodesic optimal pulses and DRAG pulses. From Fig.~\ref{fig:X90_pulses}, we notice that although $\Omega_y(t)$ components are not proportional to the first derivative of the $\Omega_x(t)$ components, it remains an interesting question for further research whether it can be written in terms of higher order derivative corrections. Finally, while this work is established on a rigorous theoretical and numerical simulation foundation, the most natural extension of this work is the practical experimental implementation of these geodesic-based pulses on physical superconducting hardware.

\begin{acknowledgments}
A.~H.~da Silva acknowledges financial support from the Postdoctoral Programme for Research Institutes (Tulanet), funded by the Finnish Government. T. Pandit and F. Cosco acknowledge financial support from the Ministry of Economic Affairs and Employment of Finland through the 300Qb quantum computer development project.
\end{acknowledgments}

\onecolumngrid
\appendix
\section{Derivation of the effective Hamiltonian}\label{appendix:calculations}

\subsection{Schrieffer-Wolff transformation}

From Eq.~(\ref{eq:Htotal}), the total Hamiltonian depends on three terms: $H_0$, given by Eq.~(\ref{eq:H0}), $H_\mathrm{int}$, given by Eq.~(\ref{eq:Hint}), and the external drive Hamiltonian $H_\mathrm{d}(t)$, given by Eq.~(\ref{eq:HAC}). The first step is to truncate the ladder operators to the first three energy levels of the qubits. The truncated $H_0$ was already written in Eq.~(\ref{eq:H0_truncated}), and for $H_\mathrm{int}$, we first write it as a sum of two parts as $H_\mathrm{int} = V_c + V_{12}$, where
\begin{align}
    V_c &= - \sum_{k=1}^2 g_k \left( a_k a_c - a_ka_c^\dagger - a_k^\dagger a_c + a_k^\dagger a_c^\dagger \right), \label{A:Vc} \\
    V_{12} &= - g_{12} \left( a_1 a_2 - a_1 a_2^\dagger - a_1^\dagger a_2 + a_1^\dagger a_2^\dagger \right). \label{A:V12}
\end{align}
As it was already mentioned, the objective is to obtain an effective Hamiltonian that lives exclusively in the subspace spanned by the two qubits, effectively removing the coupler's degrees of freedom. For that, we consider that since $\omega_c \gg \omega_k$ ($k=1,2$), the coupler interacts in the dispersive regime, such that $V_c$ can be treated as a small perturbation.

We now explicitly carry out the SW transformation for that end. Performing an SW transformation means calculating some operator $W$ such that 
\begin{align}
    \comm{W}{H_0} = - V_c,
\end{align}
and then defining the dispersive regime Hamiltonian $H_\mathrm{disp} \equiv e^W \left( H_0 + H_\mathrm{int} + H_\mathrm{d}(t) \right) e^{-W}$, and writing it as the perturbative series expansion up to second order in $V_c$ %\ahs{Note: change $O(V_c^3)$ to $O(g_k^3)$?}
as
\begin{align}\label{A:Hdisp}
    H_\mathrm{disp} = H_0 + V_{12} + H_\mathrm{d}(t) + \frac{1}{2}\comm{W}{V_c} + \comm{W}{V_{12}} + \comm{W}{H_\mathrm{d}(t)} + O(V_c^3)
\end{align}
It is direct to verify that, considering $\{\ket{\mu}\}$ to form the basis for which $H_0$ is diagonal, that is $H_0 = \sum_\mu E_\mu \ketbra{\mu}{\mu}$, 
\begin{align}
    W =  \sum_{\mu,\nu} \frac{\braket{\mu}{V_c|\nu}}{E_\mu-E_\nu} \ketbra{\mu}{\nu}
\end{align}
satisfies this condition. We adopt the explicit notation for the basis elements $\mu$:
\begin{align}
    \ket{\mu} = \ket{m_1 m_2, m_c},
\end{align}
where $m_k$, for $k=1,2,c$, ranges from $0$ to $2$. When it is convenient to show only the state of one of the qubits, we use the notation $\ket{m_q, m_c}$, by explicitly identifying the qubit with the index $q$. Notice that we can obtain the eigenenergies $E_\mu$ directly from Eq.~(\ref{eq:H0_truncated}).

Using Eq.~(\ref{A:Vc}), we have thirty-two non-zero entries in the SW operator, which are given by
\begin{align}
    \braket{1_q,0_c}{W|0_q,1_c} &= \frac{\braket{1_q,0_c}{V_c|0_q,1_c}}{E_{1_q,0_c}-E_{0_q,1_c}} = \frac{g_q}{\omega_q - \omega_c} = \frac{g_q}{\Delta_q}, \\
    \braket{1_q,1_c}{W|0_q,0_c} &= \frac{\braket{1_q,1_c}{V_c|0_q,0_c}}{E_{1_q,1_c}-E_{0_q,0_c}} = -\frac{g_q}{\omega_q + \omega_c} = -\frac{g_q}{\Sigma_q}, \\
    \braket{1_q,1_c}{W|0_q,2_c} &= \frac{\braket{1_q,1_c}{V_c|0_q,2_c}}{E_{1_q,1_c}-E_{0_q,2_c}} = \frac{\sqrt{2} g_q}{\omega_q - \omega_c - \alpha_c} = \frac{\sqrt{2} g_q}{\Delta_q - \alpha_c}, \\
    \braket{1_q,2_c}{W|0_q,1_c} &= \frac{\braket{1_q,2_c}{V_c|0_q,1_c}}{E_{1_q,2_c}-E_{0_q,1_c}} = -\frac{\sqrt{2} g_q}{\omega_q + \omega_c + \alpha_c} = -\frac{\sqrt{2} g_q}{\Sigma_q + \alpha_c}, \\
    %%%
    \braket{2_q,0_c}{W|1_q,1_c} &= \frac{\braket{2_q,0_c}{V_c|1_q,1_c}}{E_{2_q,0_c}-E_{1_q,1_c}} = \frac{\sqrt{2} g_q}{\omega_q + \alpha_q - \omega_c} = \frac{\sqrt{2} g_q}{\Delta_q + \alpha_q}, \\
    \braket{2_q,1_c}{W|1_q,0_c} &= \frac{\braket{2_q,1_c}{V_c|1_q,0_c}}{E_{2_q,1_c}-E_{1_q,0_c}} = -\frac{\sqrt{2} g_q}{\omega_q + \alpha_q + \omega_c} = -\frac{\sqrt{2} g_q}{\Sigma_q + \alpha_q}, \\
    \braket{2_q,1_c}{W|1_q,2_c} &= \frac{\braket{2_q,1_c}{V_c|1_q,2_c}}{E_{2_q,1_c}-E_{1_c,2_c}} = \frac{2 g_q}{\omega_q + \alpha_q - \omega_c - \alpha_c} = \frac{2 g_q}{\Delta_q + \alpha_q - \alpha_c}, \\
    \braket{2_q,2_c}{W|1_q,1_c} &= \frac{\braket{2_q,2_c}{V_c|1_q,1_c}}{E_{2_q,2_c}-E_{1_q,1_c}} = -\frac{2 g_q}{\omega_q + \alpha_q + \omega_c + \alpha_c} = -\frac{2 g_q}{\Sigma_q + \alpha_q + \alpha_c},
\end{align}
for $q=1,2$, and the corresponding sixteen Hermitian conjugate entries, which are given by the same values but multiplied by $-1$. Using these results, the SW operator can be written as
\begin{equation}\label{A:SW_transformator}
\begin{aligned}
    W &= \sum_{q=1}^2 \left[ \tfrac{g_q}{\Delta_q} \left( \ketbra{1_q,0_c}{0_q,1_c} - \ketbra{0_q,1_c}{1_q,0_c} \right) + \tfrac{\sqrt{2} g_q}{\Delta_q + \alpha_q} \left( \ketbra{2_q,0_c}{1_q,1_c} - \ketbra{1_q,1_c}{2_q,0_c} \right) \right. \\
    %%%
    &\quad\quad\quad\quad - \left. \tfrac{g_q}{\Sigma_q} \left( \ketbra{1_q,1_c}{0_q,0_c} - \ketbra{0_q,0_c}{1_q,1_c} \right) - \tfrac{\sqrt{2} g_q}{\Sigma_q + \alpha_q} \left( \ketbra{2_q,1_c}{1_q,0_c} - \ketbra{1_q,0_c}{2_q,1_c} \right) \right. \\
    %%%
    &\quad\quad\quad\quad + \left. \tfrac{\sqrt{2} g_q}{\Delta_q - \alpha_c} \left( \ketbra{1_q,1_c}{0_q,2_c} - \ketbra{0_q,2_c}{1_q,1_c} \right) + \tfrac{2 g_q}{\Delta_q + \alpha_q - \alpha_c} \left( \ketbra{2_q,1_c}{1_q,2_c} - \ketbra{1_q,2_c}{2_q,1_c} \right) \right. \\
    %%%
    &\quad\quad\quad\quad - \left. \tfrac{\sqrt{2} g_q}{\Sigma_q + \alpha_c} \left( \ketbra{1_q,2_c}{0_q,1_c} - \ketbra{0_q,1_c}{1_q,2_c} \right) - \tfrac{2 g_q}{\Sigma_q + \alpha_q + \alpha_c} \left( \ketbra{2_q,2_c}{1_q,1_c} - \ketbra{1_q,1_c}{2_q,2_c} \right) \right].  
\end{aligned}
\end{equation}

The next step is to explicitly calculate the three commutators that appear in Eq.~(\ref{A:Hdisp}). Instead of writing them with all terms, here we use the fact that the last step for the complete SW transformation is to apply a projection over the ground state of the coupler subspace. In other words, we obtain the effective Hamiltonian by calculating $H_\mathrm{eff} \equiv \braket{0_c}{H_\mathrm{disp} | 0_c}$, which means considering that the coupler is not effectively populated during the time evolution. Thus combining Eqs.~(\ref{A:Vc}) and~(\ref{A:SW_transformator}) we obtain
\begin{equation}\label{A:0WVc0}
\begin{aligned}
    \braket{0_c}{\comm{W}{V_c}|0_c} &= \sum_{q=1}^2 \left[ -\tfrac{2g_q^2}{\Sigma_q} \ketbra{0_q}{0_q} +  \left( \tfrac{2g_q^2}{\Delta_q} - \tfrac{4g_q^2}{\Sigma_q + \alpha_q} \right) \ketbra{1_q}{1_q} + \tfrac{4g_q^2}{\Delta_q + \alpha_q} \ketbra{2_q}{2_q} \right. \\
    &\quad\quad\quad\quad + \left. \left( \tfrac{g_q^2}{\Sigma_q} - \tfrac{g_q^2}{\Delta_q+\alpha_q} \right) \sqrt{2} \left(\ketbra{2_q}{0_q} + \ketbra{0_q}{2_q} \right) \right] \\
    &\quad + g_1g_2 \left( \tfrac{1}{\Delta_1} + \tfrac{1}{\Delta_2} - \tfrac{1}{\Sigma_1} - \tfrac{1}{\Sigma_2} \right) \left( \ketbra{10}{01} + \ketbra{01}{10} - \ketbra{11}{00} - \ketbra{00}{11} \right) \\
    &\quad + g_1g_2 \left( \tfrac{1}{\Delta_1} + \tfrac{1}{\Delta_2+\alpha_2} - \tfrac{1}{\Sigma_1} - \tfrac{1}{\Sigma_2+\alpha_2} \right) \sqrt{2} \left( \ketbra{11}{02} + \ketbra{02}{11} - \ketbra{12}{01} - \ketbra{01}{12} \right) \\
    &\quad + g_1g_2 \left( \tfrac{1}{\Delta_1+\alpha_1} + \tfrac{1}{\Delta_2} - \tfrac{1}{\Sigma_1+\alpha_1} - \tfrac{1}{\Sigma_2} \right) \sqrt{2} \left( \ketbra{20}{11} + \ketbra{11}{20} - \ketbra{21}{10} - \ketbra{10}{21} \right) \\
    &\quad + g_1g_2 \left( \tfrac{1}{\Delta_1+\alpha_1} + \tfrac{1}{\Delta_2+\alpha_2} - \tfrac{1}{\Sigma_1+\alpha_1} - \tfrac{1}{\Sigma_2+\alpha_2} \right) 2 \left( \ketbra{21}{12} + \ketbra{12}{21} - \ketbra{22}{11} - \ketbra{11}{22} \right).
\end{aligned}
\end{equation}
For the second and third commutators, notice that all terms in $V_c$ contain exactly either one operator $a_c$ or one operator $a_c^\dagger$, while $V_{12}$ and $H_\mathrm{d}(t)$ do not have any coupler operators. This means that both $\comm{W}{V_{12}}$ and $\comm{W}{H_\mathrm{d}(t)}$ have vanishing diagonal in the coupler subspace, and therefore.
\begin{align}
    \braket{0_c}{\comm{W}{V_{12}}|0_c} &= 0, \\
    \braket{0_c}{\comm{W}{H_\mathrm{d}(t)}|0_c} &= 0, 
\end{align}
Now, the only remaining terms to be calculated are
\begin{align}
    \braket{0_c}{H_0|0_c} &= \sum_{k=1}^2 \omega_k \ketbra{1_k}{1_k} + (2\omega_k + \alpha_k) \ketbra{2_k}{2_k}, \\
    \braket{0_c}{V_{12} | 0_c} &= V_{12}, \\
    \braket{0_c}{H_\mathrm{d}(t) | 0_c} &= H_\mathrm{d}(t).
\end{align}
From restricting the ladder operators to three dimensions in Eq.~(\ref{A:V12}), we can explicitly write
\begin{equation}
\begin{aligned}\label{A:V12_explicit}
    V_{12} &= g_{12} \left( \ketbra{10}{01} + \ketbra{01}{10} - \ketbra{11}{00} - \ketbra{00}{11} \right) + g_{12} \sqrt{2}\left( \ketbra{11}{02} + \ketbra{02}{11} - \ketbra{12}{01} - \ketbra{01}{12} \right) \\
    &\quad + g_{12} \sqrt{2}\left( \ketbra{20}{11} + \ketbra{11}{20} - \ketbra{21}{10} - \ketbra{10}{21} \right) + g_{12} 2\left( \ketbra{21}{12} + \ketbra{12}{21} - \ketbra{22}{11} - \ketbra{11}{22} \right).
\end{aligned}
\end{equation}
Combining all the results from Eq.~(\ref{A:0WVc0}) to Eq.~(\ref{A:V12_explicit}), with Eq.~(\ref{A:Hdisp}) yields
\begin{equation}
\begin{aligned}
    H_\mathrm{eff} &= \sum_{k=1}^2 \left[ -\tfrac{g_k^2}{\Sigma_k} \ketbra{0_k}{0_k} +  \left(\omega_k + \tfrac{g_k^2}{\Delta_k} - \tfrac{2g_k^2}{\Sigma_k + \alpha_k} \right) \ketbra{1_k}{1_k} + \left( 2\omega_k + \alpha_k + \tfrac{2g_k^2}{\Delta_k + \alpha_k} \right) \ketbra{2_k}{2_k} \right] \\
    &\quad + \sum_{k=1}^2 \tfrac{1}{2}\left( \tfrac{g_k^2}{\Sigma_k} - \tfrac{g_k^2}{\Delta_k+\alpha_k} \right) \sqrt{2} \left(\ketbra{2_k}{0_k} + \ketbra{0_k}{2_k} \right) + H_\mathrm{d}(t) \\
    &\quad + \left[ g_{12} + \tfrac{g_1g_2}{2} \left( \tfrac{1}{\Delta_1} + \tfrac{1}{\Delta_2} - \tfrac{1}{\Sigma_1} - \tfrac{1}{\Sigma_2} \right) \right] \left( \ketbra{10}{01} + \ketbra{01}{10} - \ketbra{11}{00} - \ketbra{00}{11} \right) \\
    &\quad + \left[ g_{12} + \tfrac{g_1g_2}{2} \left( \tfrac{1}{\Delta_1} + \tfrac{1}{\Delta_2+\alpha_2} - \tfrac{1}{\Sigma_1} - \tfrac{1}{\Sigma_2+\alpha_2} \right) \right] \sqrt{2} \left( \ketbra{11}{02} + \ketbra{02}{11} - \ketbra{12}{01} - \ketbra{01}{12} \right) \\
    &\quad + \left[ g_{12} + \tfrac{g_1g_2}{2} \left( \tfrac{1}{\Delta_1+\alpha_1} + \tfrac{1}{\Delta_2} - \tfrac{1}{\Sigma_1+\alpha_1} - \tfrac{1}{\Sigma_2} \right) \right] \sqrt{2} \left( \ketbra{20}{11} + \ketbra{11}{20} - \ketbra{21}{10} - \ketbra{10}{21} \right) \\
    &\quad + \left[ g_{12} + \tfrac{g_1g_2}{2} \left( \tfrac{1}{\Delta_1+\alpha_1} + \tfrac{1}{\Delta_2+\alpha_2} - \tfrac{1}{\Sigma_1+\alpha_1} - \tfrac{1}{\Sigma_2+\alpha_2} \right) \right] 2 \left( \ketbra{21}{12} + \ketbra{12}{21} - \ketbra{22}{11} - \ketbra{11}{22} \right).
\end{aligned}
\end{equation}
Notice that there is a negative correction to the ground state energy. Just as initially we considered the energy of the ground state to be null in $H_0$, we can displace the energy reference by simply adding $g_k^2/\Sigma_k$ to the other two diagonal terms. We can also identify the four effective coupling constants
\begin{align}
    \Gamma_{\downarrow \downarrow} &\equiv g_{12} + \tfrac{g_1g_2}{2} \left( \tfrac{1}{\Delta_1} + \tfrac{1}{\Delta_2} - \tfrac{1}{\Sigma_1} - \tfrac{1}{\Sigma_2} \right), \\
    \Gamma_{\downarrow \uparrow} &\equiv g_{12} + \tfrac{g_1g_2}{2} \left( \tfrac{1}{\Delta_1} + \tfrac{1}{\Delta_2+\alpha_2} - \tfrac{1}{\Sigma_1} - \tfrac{1}{\Sigma_2+\alpha_2} \right), \\
    \Gamma_{\uparrow \downarrow} &\equiv g_{12} + \tfrac{g_1g_2}{2} \left( \tfrac{1}{\Delta_1+\alpha_1} + \tfrac{1}{\Delta_2} - \tfrac{1}{\Sigma_1+\alpha_1} - \tfrac{1}{\Sigma_2} \right), \\
    \Gamma_{\uparrow \uparrow} &\equiv g_{12} + \tfrac{g_1g_2}{2} \left( \tfrac{1}{\Delta_1+\alpha_1} + \tfrac{1}{\Delta_2+\alpha_2} - \tfrac{1}{\Sigma_1+\alpha_1} - \tfrac{1}{\Sigma_2+\alpha_2} \right),
\end{align}
which appear in Eq.~(\ref{eq:Heff}), and also the amplitudes associated with two-photons exchanges
\begin{align}
    \Gamma_k^{(\mathrm{tp})} \equiv \frac{1}{2}\left( \frac{g_k^2}{\Sigma_k} - \frac{g_k^2}{\Delta_k+\alpha_k} \right).
\end{align}
With these definitions, the effective Hamiltonian becomes
\begin{equation}
\begin{aligned}
    H_\mathrm{eff} &= \sum_{k=1}^2 \left[ \left(\omega_k + \tfrac{g_k^2}{\Delta_k} + \tfrac{g_k^2}{\Sigma_k} - \tfrac{2g_k^2}{\Sigma_k + \alpha_k} \right) \ketbra{1_k}{1_k} + \left( 2\omega_k + \alpha_k + \tfrac{g_k^2}{\Sigma_k} + \tfrac{2g_k^2}{\Delta_k + \alpha_k} \right) \ketbra{2_k}{2_k} \right] \\
    &\quad + \sum_{k=1}^2 \Gamma_k^{(\mathrm{tp})} \sqrt{2} \left(\ketbra{2_k}{0_k} + \ketbra{0_k}{2_k} \right) + H_\mathrm{d}(t) \\
    &\quad + \Gamma_{\downarrow \downarrow} \left( \ketbra{10}{01} + \ketbra{01}{10} - \ketbra{11}{00} - \ketbra{00}{11} \right) + \Gamma_{\downarrow \uparrow} \sqrt{2} \left( \ketbra{11}{02} + \ketbra{02}{11} - \ketbra{12}{01} - \ketbra{01}{12} \right) \\
    &\quad + \Gamma_{\uparrow \downarrow} \sqrt{2} \left( \ketbra{20}{11} + \ketbra{11}{20} - \ketbra{21}{10} - \ketbra{10}{21} \right) + \Gamma_{\uparrow \uparrow} 2 \left( \ketbra{21}{12} + \ketbra{12}{21} - \ketbra{22}{11} - \ketbra{11}{22} \right).
\end{aligned}
\end{equation}
Finally, we can identify the dressed frequencies and anharmonicities, shown in Eqs.~(\ref{eq:omegaktilde}) and~(\ref{eq:alphaktilde}),
\begin{align}
    \widetilde{\omega}_k &\equiv \omega_k + \frac{g_k^2}{\Delta_k} + \frac{g_k^2}{\Sigma_k} - \frac{2g_k^2}{\Sigma_k + \alpha_k}, \\
    \widetilde{\alpha}_k &\equiv \alpha_k - \frac{2g_k^2}{\Delta_k} + \frac{2g_k^2}{\Delta_k+\alpha_k} - \frac{g_k^2}{\Sigma_k} + \frac{4g_k^2}{\Sigma_k+\alpha_k},
\end{align}
and rewrite the effective Hamiltonian as
\begin{equation}
\begin{aligned}\label{A:Heff_original}
    H_\mathrm{eff} &= \sum_{k=1}^2 \left( \widetilde{\omega}_k \ketbra{1_k}{1_k} + \left( 2\widetilde{\omega}_k + \widetilde{\alpha}_k \right) \ketbra{2_k}{2_k} \right) + \sum_{k=1}^2 \Gamma_k^{(\mathrm{tp})} \sqrt{2} \left(\ketbra{2_k}{0_k} + \ketbra{0_k}{2_k} \right) + H_\mathrm{d}(t) \\
    &\quad + \Gamma_{\downarrow \downarrow} \left( \ketbra{10}{01} + \ketbra{01}{10} - \ketbra{11}{00} - \ketbra{00}{11} \right) + \Gamma_{\downarrow \uparrow} \sqrt{2} \left( \ketbra{11}{02} + \ketbra{02}{11} - \ketbra{12}{01} - \ketbra{01}{12} \right) \\
    &\quad + \Gamma_{\uparrow \downarrow} \sqrt{2} \left( \ketbra{20}{11} + \ketbra{11}{20} - \ketbra{21}{10} - \ketbra{10}{21} \right) + \Gamma_{\uparrow \uparrow} 2 \left( \ketbra{21}{12} + \ketbra{12}{21} - \ketbra{22}{11} - \ketbra{11}{22} \right).
\end{aligned}
\end{equation}

\subsection{Interaction picture}

Equation~(\ref{A:Heff_original}) is a Hamiltonian involving only the two-qubits. The only remaining step is to transform it into the frame co-rotating with the two qubits' dressed frequencies. This is done by defining the unitary operator $U_\mathrm{ho}(t) \equiv \exp(-i H_\mathrm{ho} t)$, where
\begin{align}
    H_\mathrm{ho} \equiv \sum_{k=1}^2 \left[ \widetilde{\omega}_k \ketbra{1_k}{1_k} + 2\widetilde{\omega}_k \ketbra{2_k}{2_k} \right]
\end{align}
can be thought of as the ``harmonic oscilator'' part of $H_0$. Here we use an abuse of notation and denote the effective Hamiltonian in the interaction picture with the same name but by explicitly writing the time dependency, as
\begin{align}\label{A:interaction_picture_transform}
    H_\mathrm{eff}(t) \equiv U_\mathrm{ho}^\dagger(t) H_\mathrm{eff} U_\mathrm{ho}(t) - H_\mathrm{ho}.
\end{align}
We will denote the control Hamiltonian in the interaction picture as $H_\mathrm{c}(t) \equiv U_\mathrm{ho}^\dagger(t) H_\mathrm{d}(t) U_\mathrm{ho}(t)$. Then, using Eqs.~(\ref{A:Heff_original}) and~(\ref{A:interaction_picture_transform}), we obtain
\begin{equation}\label{A:Heff_almost}
\begin{aligned}
    H_\mathrm{eff}(t) &= \sum_{k=1}^2 \widetilde{\alpha}_k \ketbra{2_k}{2_k} + \sum_{k=1}^2 \Gamma_k^{(\mathrm{tp})} \sqrt{2} \left( e^{i 2\widetilde{\omega}_k t} \ketbra{2_k}{0_k} + e^{-i 2\widetilde{\omega}_k t} \ketbra{0_k}{2_k} \right) + H_\mathrm{c}(t) \\
    &\quad + \Gamma_{\downarrow \downarrow} \left( e^{i \widetilde{\omega}_1 t} e^{-i \widetilde{\omega}_2 t} \ketbra{10}{01} + e^{-i \widetilde{\omega}_1 t} e^{i \widetilde{\omega}_2 t} \ketbra{01}{10} - e^{i \widetilde{\omega}_1 t} e^{i \widetilde{\omega}_2 t} \ketbra{11}{00} - e^{-i \widetilde{\omega}_1 t} e^{-i \widetilde{\omega}_2 t} \ketbra{00}{11} \right) \\
    &\quad + \Gamma_{\downarrow \uparrow} \sqrt{2} \left( e^{i \widetilde{\omega}_1 t} e^{-i \widetilde{\omega}_2 t} \ketbra{11}{02} + e^{-i \widetilde{\omega}_1 t} e^{i \widetilde{\omega}_2 t} \ketbra{02}{11} - e^{i \widetilde{\omega}_1 t} e^{i \widetilde{\omega}_2 t} \ketbra{12}{01} - e^{-i \widetilde{\omega}_1 t} e^{-i \widetilde{\omega}_2 t} \ketbra{01}{12} \right) \\
    &\quad + \Gamma_{\uparrow \downarrow} \sqrt{2} \left( e^{i \widetilde{\omega}_1 t} e^{-i \widetilde{\omega}_2 t} \ketbra{20}{11} + e^{-i \widetilde{\omega}_1 t} e^{i \widetilde{\omega}_2 t} \ketbra{11}{20} - e^{i \widetilde{\omega}_1 t} e^{i \widetilde{\omega}_2 t} \ketbra{21}{10} - e^{-i \widetilde{\omega}_1 t} e^{-i \widetilde{\omega}_2 t} \ketbra{10}{21} \right) \\
    &\quad + \Gamma_{\uparrow \uparrow} 2 \left( e^{i \widetilde{\omega}_1 t} e^{-i \widetilde{\omega}_2 t} \ketbra{21}{12} + e^{-i \widetilde{\omega}_1 t} e^{i \widetilde{\omega}_2 t} \ketbra{12}{21} - e^{i \widetilde{\omega}_1 t} e^{i \widetilde{\omega}_2 t} \ketbra{22}{11} - e^{-i \widetilde{\omega}_1 t} e^{-i \widetilde{\omega}_2 t} \ketbra{11}{22} \right).
\end{aligned}
\end{equation}

Let us now specifically focus on $H_\mathrm{c}(t)$. The drive Hamiltonian $H_\mathrm{d}(t)$ is shown in Eq.~(\ref{eq:Hd_original}) with the ladder operators truncated to three dimensions. Writing the functions $V_{\mathrm{d},k}(t)$ explicitly, using Eq.~(\ref{eq:Vdk_original}), we obtain
\begin{equation}
\begin{aligned}
    H_\mathrm{c}(t) &= \sum_{k=1}^2 -i c_k \Big[ V_{I,k}(t) \cos((\omega_{\mathrm{d},k} t + \phi_k)) + V_{Q,k}(t) \sin(\omega_{\mathrm{d},k} t + \phi_k) \Big] \\
    &\quad\quad\quad\quad\quad \times \left[ e^{-i \widetilde{\omega}_k t} \ketbra{0_k}{1_k} - e^{i \widetilde{\omega}_k t}\ketbra{1_k}{0_k} + \sqrt{2} \left( e^{-i \widetilde{\omega}_k t}\ketbra{1_k}{2_k} - e^{i \widetilde{\omega}_k t}\ketbra{2_k}{1_k} \right) \right].
\end{aligned}
\end{equation}
Rewriting the trigonometric functions in exponential form yields
\begin{equation}
\begin{aligned}
    H_\mathrm{c}(t) &= \sum_{k=1}^2 -i c_k \left( \frac{V_{I,k}(t) - i V_{Q,k}(t)}{2} \right) e^{i\phi_k} \Big[ e^{i \omega_{\mathrm{d},k} t} e^{- i \widetilde{\omega}_k t} \ketbra{0_k}{1_k} - e^{i \omega_{\mathrm{d},k} t} e^{i \widetilde{\omega}_k t}\ketbra{1_k}{0_k} \\ 
    &\quad \hspace{5.75cm} + \sqrt{2} \left( e^{i \omega_{\mathrm{d},k} t} e^{-i \widetilde{\omega}_k t}\ketbra{1_k}{2_k} - e^{i \omega_{\mathrm{d},k} t} e^{i \widetilde{\omega}_k t}\ketbra{2_k}{1_k} \right) \Big] \\
    &\phantom{\sum_{k=1}^2} \quad -i c_k \left( \frac{V_{I,k}(t) + i V_{Q,k}(t)}{2} \right) e^{-i\phi_k} \Big[ e^{-i \omega_{\mathrm{d},k} t} e^{-i \widetilde{\omega}_k t} \ketbra{0_k}{1_k} - e^{-i \omega_{\mathrm{d},k} t} e^{i \widetilde{\omega}_k t}\ketbra{1_k}{0_k} \\
    &\quad\quad \hspace{5.75cm} + \sqrt{2} \left( e^{-i \omega_{\mathrm{d},k} t} e^{-i \widetilde{\omega}_k t}\ketbra{1_k}{2_k} - e^{-i \omega_{\mathrm{d},k} t} e^{i \widetilde{\omega}_k t}\ketbra{2_k}{1_k} \right) \Big].
\end{aligned}
\end{equation}
At this point we can define the detunings between the $k$-th qubit and its drive as $\Delta_k \equiv \omega_{\mathrm{d},k} - \widetilde{\omega}_k$, and considering the external drive will not be too off-resonant, we can apply the rotating wave approximation (RWA), effectively neglecting terms oscillating fast with frequency $\omega_{\mathrm{d},k} + \widetilde{\omega}_k$. This yields
\begin{equation}
\begin{aligned}
    H_\mathrm{c}(t) &\approx \sum_{k=1}^2 -i c_k \left( \frac{V_{I,k}(t) - i V_{Q,k}(t)}{2} \right) e^{i (\Delta_k t + \phi_k)} \Big[ \ketbra{0_k}{1_k} + \sqrt{2}\ketbra{1_k}{2_k} \Big] \\
    &\phantom{\sum_{k=1}^2} \quad +i c_k \left( \frac{V_{I,k}(t) + i V_{Q,k}(t)}{2} \right) e^{-i(\Delta_k t + \phi_k)} \Big[ \ketbra{1_k}{0_k} + \sqrt{2}\ketbra{2_k}{1_k} \Big].
\end{aligned}
\end{equation}
Rewriting the exponentials back in trigonometric form and defining $\theta_k(t) \equiv \Delta_k t + \phi_k$ gives
\begin{equation}
\begin{aligned}\label{A:Hc_almost}
    H_\mathrm{c}(t) &= \sum_{k=1}^2 \frac{c_k}{2} \left[ V_{I,k}(t) \sin{\theta_k(t)} - V_{Q,k}(t) \cos{\theta_k(t)} \right] \left[ \ketbra{0_k}{1_k} + \ketbra{1_k}{0_k} + \sqrt{2}\left( \ketbra{1_k}{2_k} + \ketbra{2_k}{1_k} \right) \right] \\
    &\quad \phantom{\sum_{k=1}^2} + \frac{c_k}{2} \left[ V_{I,k}(t) \cos{\theta_k(t)} + V_{Q,k}(t) \sin{\theta_k(t)} \right] \left[ -i\ketbra{0_k}{1_k} + i \ketbra{1_k}{0_k} + \sqrt{2}\left( -i \ketbra{1_k}{2_k} + i \ketbra{2_k}{1_k} \right) \right].
\end{aligned}
\end{equation}
Here we can identify the matrices
\begin{align}
    \lambda_{x,k} &\equiv \frac{1}{\sqrt{3}} \left[ \ketbra{0_k}{1_k} + \ketbra{1_k}{0_k} + \sqrt{2}\left( \ketbra{1_k}{2_k} + \ketbra{2_k}{1_k} \right) \right], \\
    \lambda_{y,k} &\equiv \frac{1}{\sqrt{3}} \left[ -i\ketbra{0_k}{1_k} + i \ketbra{1_k}{0_k} + \sqrt{2}\left( -i \ketbra{1_k}{2_k} + i \ketbra{2_k}{1_k} \right) \right],
\end{align}
defined in Eq.~(\ref{eq:lambx_and_lamby}), and the Rabi frequencies
\begin{align}
    \Omega_{x,k}(t) &\equiv \frac{\sqrt{3}}{2}c_k \left[ V_{I,k}(t) \sin{\theta_k(t)} - V_{Q,k}(t) \cos{\theta_k(t)} \right], \\
    \Omega_{y,k}(t) &\equiv \frac{\sqrt{3}}{2}c_k \left[ V_{I,k}(t) \cos{\theta_k(t)} + V_{Q,k}(t) \sin{\theta_k(t)} \right],
\end{align}
defined in Eq.~(\ref{eq:Rabi}). Substituting these back into Eq.~(\ref{A:Hc_almost}) results in the interaction picture control Hamiltonian shown in Eq.~(\ref{eq:Hd}).

Returning to Eq.~(\ref{A:Heff_almost}), the last step is to also perform a RWA by considering that the two qubits are not too off-resonant from each other. For that, we identity the detuning between the perturbed frequencies of the two qubits $\widetilde{\Delta} \equiv \widetilde{\omega}_2 - \widetilde{\omega}_1$, and then, neglect all terms oscillating with frequency $\widetilde{\omega}_2 + \widetilde{\omega}_1$ and $2 \widetilde{\omega}_k$. Doing this approximation and writing $H_\mathrm{c}(t)$ explicitly finally yields the effective Hamiltonian used in this work, shown in Eq.~(\ref{eq:Heff}):
\begin{equation}
\begin{aligned}
    H_\mathrm{eff}(t) &= \sum_{k=1}^2 \widetilde{\alpha}_k \ketbra{2_k}{2_k} + \sum_{k=1}^2 \left[ \Omega_{x,k}(t) \lambda_{x,k} + \Omega_{y,k}(t) \lambda_{y,k}, \right] \\
    &\quad + e^{-i \widetilde{\Delta} t} \left[ \Gamma_{\downarrow\downarrow} \ketbra{10}{01} + \sqrt{2} \left( \Gamma_{\downarrow\uparrow} \ketbra{11}{02} + \Gamma_{\uparrow\downarrow} \ketbra{20}{11} \right) + 2 \Gamma_{\uparrow \uparrow} \ketbra{21}{12} \right] \\
    &\quad + e^{i\widetilde{\Delta}t} \left[ \Gamma_{\downarrow\downarrow} \ketbra{01}{10} + \sqrt{2} \left( \Gamma_{\downarrow\uparrow} \ketbra{02}{11} + \Gamma_{\uparrow\downarrow} \ketbra{11}{20} \right) + 2 \Gamma_{\uparrow \uparrow} \ketbra{12}{21} \right].
\end{aligned}
\end{equation}

\twocolumngrid
\bibliography{refs}

\end{document}